\documentclass[12pt]{iopart}
\usepackage{graphics}
\usepackage{bm}
\usepackage{amsfonts}
\usepackage{multicol}
\usepackage{cite}
\begin{document}
\input{psfig}
\bibliographystyle{unsrt}

\title{Radiative Transport Equation in Rotated Reference Frames.}

\author{George Panasyuk, John C. Schotland}

\address{Department of Bioengineering, University of Pennsylvania,
  Philadelphia, PA 19104}
 
\author{Vadim A. Markel\footnote[1]{To
whom correspondence should be addressed (vmarkel@mail.med.upenn.edu)}}

\address{Department of Radiology, University of Pennsylvania,
  Philadelphia, PA 19104}

\begin{abstract}
  A novel method for solving the linear radiative transport equation
  (RTE) in a three-dimensional homogeneous medium is proposed and
  illustrated with numerical examples. The method can be used with an
  arbitrary phase function $A(\hat{\bf s},\hat{\bf s}^{\prime})$ with
  the constraint that it depends only on the angle between the angular
  variables $\hat{\bf s}$ and $\hat{\bf s}^{\prime}$. This corresponds
  to spherically symmetric (on average) random medium constituents.
  Boundary conditions are considered in the slab and half-space
  geometries. The approach developed in this paper is {\it spectral}.
  It allows for the expansion of the solution to the RTE in terms of
  analytical functions of angular and spatial variables to relatively
  high orders. The coefficients of this expansion must be computed
  numerically. However, the computational complexity of this task is
  much smaller than in the standard method of spherical harmonics. The
  solutions obtained are especially convenient for solving inverse
  problems associated with radiative transfer.
\end{abstract}

\submitto{\JPA}
\maketitle

\section{Introduction}
\label{sec:intro}

\subsection{Background}
\label{subsec:background}

The contemporary mesoscopic theoretical description of multiple
scattering of waves in random media is most often based on the linear
radiative transport equation (RTE)~\cite{rossum_99_1}. Unfortunately,
the RTE is notoriously difficult to solve, even in the case of
constant absorption and scattering coefficients. The known analytical
solutions are few and of little practical importance. Yet, there is a
growing need for accurate and computationally efficient solutions to
the RTE in many fields of applied and fundamental science. For
example, in optical tomography of biological
tissues~\cite{boas_01_1,gibson_05_1}, the use of the RTE is frequently
required to accurately describe propagation of multiply scattered
light. This is especially true in close proximity to sources or
boundaries~\cite{amic_96_1}, or in regions with high absorption and
low scattering~\cite{firbank_96_1,hielscher_98_1}.  Accordingly,
significant effort has been devoted to developing and refining
efficient approximate and numerical methods for solving the RTE. In
particular, recently explored approaches have been based on the
discrete ordinate method~\cite{kim_03_1,kim_04_1,ren_04_1}, cumulant
expansion~\cite{cai_00_1,xu_02_1}, modifications of the Ambarzumian's
method~\cite{mueller_02_3,mueller_02_4}, and different levels of the
$P_L$ approximation~\cite{jiang_99_1,hull_01_1}. Algorithms for
inversion of the RTE have also been
proposed~\cite{klose_02_1,klose_02_2,abdoulaev_03_1,cai_03_1}.

The discrete ordinate method (see~\cite{thomas_book_99} for a detailed
description) is, perhaps, the most common approach due to its
simplicity and generality. An alternative to the discrete ordinate
method is the method of spherical harmonics, often referred to as the
$P_L$ approximation, in cases with special symmetry. This approach has
the advantage of expressing the angular dependence of the specific
intensity in a basis of analytical functions~\footnote[2]{Here the
  term ``analytical'' means ``expressed in terms of well-characterized
  functions through explicit formulas'', not necessarily analytic in
  the Cauchy-Riemann sense.} rather than in the completely local basis
of discrete ordinates. In particular, in the case of cylindrical
symmetry (one-dimensional propagation), a very effective solution
based on a contuned fraction expansion can be
obtained~\cite{boguna_00_1}.  However, when no special symmetry is
present in the problem, the method of spherical harmonics can be
carried out in practice only to very low orders. In a recent
paper~\cite{markel_04_1}, we have suggested a modification of the
standard method of spherical harmonics. The modification is based on
expanding the angular part of each Fourier component of the specific
intensity in the basis of spherical functions defined in a reference
frame whose $z$-axis is aligned with the direction of the Fourier wave
vector. This approach resulted in significant mathematical
simplifications and was referred to as {\em the modified method of
  spherical harmonics} in~\cite{markel_04_1}. Here we find it more
appropriate to call it {\em the method of rotated reference frames}
(MRRF).

In Ref.~\cite{markel_04_1}, the derivation of the RTE Green's function
by the MRRF was only briefly sketched in infinite-medium and numerical
examples were limited to a few simple cases with spherical symmetry.
Here we give the full mathematical details of the derivation and
discuss the mathematical properties of the solutions obtained, derive
plane-wave decomposition of the Green's function, and generalize the
MRRF to the case of planar boundaries. We also provide extensive
numerical examples for cases with no special symmetry. The paper is
organized as follows. In Section~\ref{subsec:RTE}, we introduce the
RTE and basic notations, and explain why the use of rotated reference
frames is beneficial. In Section~\ref{sec:rot_frames} we define
spherical functions in rotated reference frames. In
Section~\ref{sec:theory} we apply the MRRF to the derivation of the
Green's function.  In particular, the Green's function in the Fourier
representation is given in Section~\ref{subsec:Green_Fur}.
Mathematical properties of the solutions are discussed in
Section~\ref{subsec:math_prop}.  Different representations for the
Green's function in real space are given in
Section~\ref{subsec:Green_RS}. A plane-wave decomposition of the
Green's function is derived in Section~\ref{subsec:pw_dec}. In
Section~\ref{subsubsec:plane_eva}, we introduce evanescent modes of
the homogeneous RTE. These modes are important mathematical constructs
which can be used for solving the RTE in the presence of planar
boundaries, as is shown in Section~\ref{subsec:boundaries}.
Section~\ref{sec:num} contains numerical examples of applying the MRRF
to calculating the Green's function in infinite space. Finally,
Section~\ref{sec:summ_disc} contains a discussion.

\subsection{RTE and the conventional method of spherical harmonics}
\label{subsec:RTE}

The RTE describes the propagation of the {\em specific intensity}
$I({\bf r},\hat{\bf s})$, at the spatial point ${\bf r}$ and flowing
in the direction specified by the unit vector $\hat{\bf s}$, in a
medium characterized by absorption and scattering coefficients $\mu_a$
and $\mu_s$, and has the form

\begin{equation}
\label{RTE}
\hat{\bf s}\cdot \nabla I + (\mu_a + \mu_s)I = \mu_s A I +
\varepsilon \ .
\end{equation}

\noindent
Here $\varepsilon = \varepsilon({\bf r},\hat{\bf s})$ is the source and
$A$ is the scattering operator defined by

\begin{equation}
\label{A_oper_def}
A I({\bf r},\hat{\bf s}) = \int A(\hat{\bf s},\hat{\bf
  s}^{\prime}) I({\bf r},\hat{\bf s}^{\prime}) d^2 \hat{\bf
  s}^{\prime} \ .
\end{equation}

\noindent
The phase function $A(\hat{\bf s},\hat{\bf s}^{\prime})$ is normalized
according to the condition $\int A(\hat{\bf s},\hat{\bf s}^{\prime})
d^2\hat{\bf s}^{\prime} = 1$. We also assume that it depends only on
the angle between $\hat{\bf s}$ and $\hat{\bf s}^{\prime}$:
$A(\hat{\bf s},\hat{\bf s}^{\prime}) = f(\hat{\bf s}\cdot\hat{\bf
  s}^{\prime})$.  This fundamental assumption is often used and
corresponds to scattering by spherically symmetric particles.

In the conventional method of spherical harmonics, all angle-dependent
quantities are expanded in the basis of spherical harmonics defined in
the laboratory frame~\cite{case_67_book}:

\begin{eqnarray}
&& I({\bf r},\hat{\bf s}) = \sum_{lm} I_{lm}({\bf r})Y_{lm}(\hat{\bf
  s}) \ , \label{I_lm_case} \\ 
&& \varepsilon({\bf r},\hat{\bf s}) = \sum_{lm} \varepsilon_{lm}({\bf
  r})Y_{lm}(\hat{\bf s})  \label{eps_lm_case} \ , \\ 
&& A(\hat{\bf s},\hat{\bf s}^{\prime}) = \sum_{lm} A_l Y_{lm}(\hat{\bf
  s}) Y_{lm}^*(\hat{\bf s}^{\prime})  \label{A_l_case} \ .
\end{eqnarray}

\noindent
In particular, truncating the above series at $l=1$ leads to the
well-known diffusion approximation to the RTE~\cite{case_67_book}. In
the more general case, substituting expansions
(\ref{I_lm_case})-(\ref{A_l_case}) into the RTE~(\ref{RTE}),
multiplying the resulting equations by
$Y_{l^{\prime}m^{\prime}}^*(\hat{\bf s})$ and integrating over
$\hat{\bf s}$ leads to the following system of equations for
$I_{lm}({\bf r})$:

\begin{equation}
\label{expansion_case}
\sum_{l^{\prime}m^{\prime}} \left[
R^{(x)}_{lm,l^{\prime}m^{\prime}}{ {\partial I_{l^{\prime}m^{\prime}}}
    \over {\partial x} } + 
R^{(y)}_{lm,l^{\prime}m^{\prime}}{ {\partial I_{l^{\prime}m^{\prime}}}
    \over {\partial y} } + 
R^{(z)}_{lm,l^{\prime}m^{\prime}}{ {\partial I_{l^{\prime}m^{\prime}}}
    \over {\partial z} } \right] + \sigma_l I_{lm} = \varepsilon_{lm}
\ ,
\end{equation}

\noindent 
where $R^{(\alpha)}=\int s_\alpha Y_{lm}^*(\hat{\bf s})
Y_{l^{\prime}m^{\prime}}(\hat{\bf s})d^2\hat{s}$ ($\alpha=x,y,z$)
are matrices whose explicit form is given in Ref.~\cite{case_67_book}
and 

\begin{equation}
\label{sigma_l_def}
\sigma_l = \mu_a + \mu_s(1 - A_l) \ .
\end{equation}

\noindent
This system of partial differential equations must be solved for
$l=0,1,2,\ldots , l_{\rm max}$ and $m=-l,\ldots, l$, where $l_{\rm
  max}$ is the truncation order of the expansion
(\ref{I_lm_case})-(\ref{A_l_case}).

In a classic text, Case and Zweifel wrote concerning the system of
equations (\ref{expansion_case}): ``This rather awe-inspiring set of
equations \ldots has perhaps only academic interest''
(Ref.~\cite{case_67_book}, p. 219). We note that the root of the
difficulty is not that the matrices $R^{(\alpha)}$ are dense (in fact,
they only couple coefficients with $m^{\prime}=m$, $l^{\prime}=l\pm 1$
for $\alpha=z$ and $m^{\prime}=m\pm 1$, $l^{\prime}=l\pm 1$ for
$\alpha=x,y$) or non-commuting (in fact, it is easy to verify that all
$R^{(\alpha)}$ commute). The difficulty is that these matrices operate
on the spatial derivatives of $I_{lm}$ taken along different
directions. Thus, by viewing the set of three matrices $R^{(\alpha)}$
as a three-dimensional vector of matrices ${\bf R}$, and using the
Fourier representation for $I_{lm}$, we can rewrite the term in the
square bracket of (\ref{expansion_case}) as $i{\bf k} \cdot {\bf
  R}_{lm,l^{\prime}m^{\prime}} I_{l^{\prime}m^{\prime}}({\bf k})$. It
can be seen that the matrix ${\bf k}\cdot{\bf R}$ depends explicitly
on the direction and length of ${\bf k}$. (See similar formulation in
Ref.~\cite{barrett_book_04}.)

The method of rotated reference frames (MRRF), similar to the
conventional method of spherical harmonics, does not lead to the
separation of spatial and angular variables which is impossible for
the RTE. However, by choosing a different ${\bf k}$-dependent angular
basis, we replace the dot product ${\bf k}\cdot{\bf R}$ by an
expression of the type $k R$, where $k=\vert {\bf k}\vert$ is a scalar
and $R$ is a single ${\bf k}$-independent block-diagonal matrix. It is
shown below that, given generalized eigenvectors and eigenvalues of
$R$ which must be computed numerically, the solution can be obtained
in terms of analytical functions of spatial and angular
variables.~\footnote[3]{In principle, it should be also possible to
  use the fact that all matrices $R^{(\alpha)}$ in
  (\ref{expansion_case}) commute and, hence, have the same set of
  eigenvectors, to solve (\ref{expansion_case}) by diagonalizing just
  one ${\bf k}$-independent matrix and analytically inverting the
  Fourier transform, and thus avoid the use of rotated reference
  frames.  This approach has some advantages and difficulties
  associated with it, and to the best of our knowledge, has not been
  explored so far.  If successful, it should lead to the same
  solutions as described below.}

\section{Spherical functions in rotated frames}
\label{sec:rot_frames}

\begin{figure}
\centerline{\psfig{file=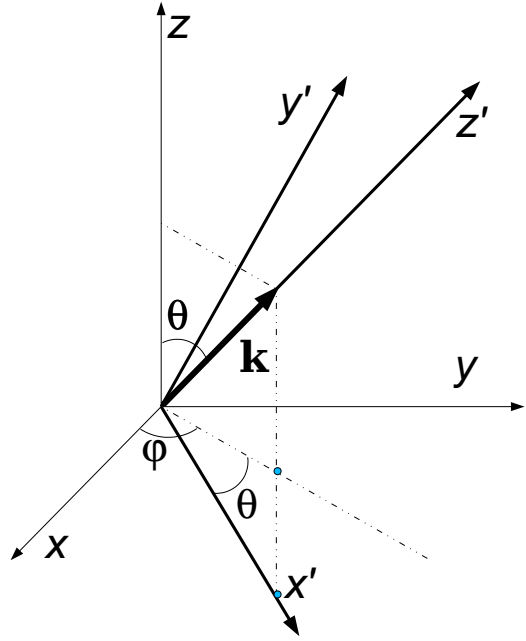,width=4in,bbllx=0bp,bblly=90bp,bburx=100bp,bbury=200bp,clip=t}}
\caption{Illustration of the rotated reference frame}
\label{fig:ref_frame}
\end{figure}

The ordinary spherical harmonics $Y_{lm}(\theta,\varphi)$ are
functions of two polar angles in a fixed (laboratory) reference frame.
Equivalently, we can view them as functions of a unit vector,
$\hat{\bf s}$. In this case, $\theta$ and $\varphi$ are the polar
angles of $\hat{\bf s}$ in the laboratory frame. More generally, both
the orientation of the reference frame and the direction of $\hat{\bf
  s}$ can vary. We will need to define spherical functions of a unit
vector $\hat{\bf s}$ in a reference frame whose $z$-axis coincides
with the direction of a given unit vector $\hat{\bf k}$. Obviously,
there are infinitely many such reference frames. To define one
uniquely, it is sufficient to consider a rotation of the laboratory
frame with the following three Euler angles: $\alpha=\varphi_{\hat{\bf
    k}}$, $\beta=\theta_{\hat{\bf k}}$ and $\gamma=0$, where
$\theta_{\hat{\bf k}}$ and $\varphi_{\hat{\bf k}}$ are the polar
angles of $\hat{\bf k}$ in the laboratory frame. The transformation
from the laboratory frame $(x,y,z)$ to the rotated frame
$(x^{\prime},y^{\prime},z^{\prime})$ is illustrated in
Fig.~\ref{fig:ref_frame}. We denote spherical functions of $\hat{\bf
  s}$ in the reference frame defined by the above transformation by
$Y_{lm}(\hat{\bf s};\hat{\bf k})$. They can be expressed as linear
combinations of the spherical functions defined in the original
(laboratory) frame according to

\begin{equation}
\label{D_l}
Y_{lm}(\hat{\bf s};\hat{\bf k}) = Y_{lm}(\hat{\bf s};\hat{\bf
  z}^{\prime}) = \sum_{m^{\prime}=-l}^l 
D_{m^{\prime}m}^l(\varphi_{\hat{\bf k}},\theta_{\hat{\bf k}},0)
Y_{lm^{\prime}}(\hat{\bf s};\hat{\bf z}) \ ,
\end{equation}

\noindent
where 

\begin{equation}
\label{dmll}
D_{mm^{\prime}}^l(\alpha,\beta,\gamma)=\exp(-im\alpha)
d_{mm^{\prime}}^l (\beta)\exp(-im^{\prime}\gamma)
\end{equation}

\noindent
are the Wigner D-functions; the explicit form of $d_{mm^{\prime}}^l
(\beta)$ is given, for example, in Ref.~\cite{varshalovich_book_88}.

It is important to note that the expansion of the scattering kernel
into the spherical functions $Y_{lm}(\hat{\bf s};\hat{\bf k})$ is
independent of the direction of $\hat{\bf k}$:

\begin{equation}
  \label{A_l_rotated} 
A(\hat{\bf s},\hat{\bf s}^{\prime}) = \sum_{lm} A_l Y_{lm}(\hat{\bf
  s};\hat{\bf k}) Y_{lm}^*(\hat{\bf s}^{\prime};\hat{\bf k}) \ .
\end{equation}

\noindent
Here the expansion coefficients $A_l$ are independent of $\hat{\bf k}$
and are the same as in (\ref{A_l_case}). This fact follows from the
rotational invariance of the scalar product.

\section{Theory} 
\label{sec:theory}

\subsection{Green's function in the Fourier representation}
\label{subsec:Green_Fur}

By definition, the Green's function $G({\bf r},\hat{\bf s};{\bf
  r}_0,\hat{\bf s}_0)$ satisfies RTE (\ref{RTE}) with the delta-source
$\varepsilon = \delta({\bf r}-{\bf r}_0) \delta(\hat{\bf
  s}-\hat{\bf s}_0)$. We will refer to ${\bf r}_0, \ \hat{\bf s}_0$
and ${\bf r}, \ \hat{\bf s}$ as the location and direction of the
source and detector, respectively. In infinite isotropic space, the
Green's function can be written in the following general form:

\begin{eqnarray}
&& G({\bf r},\hat{\bf s};{\bf r}_0,\hat{\bf s}_0) = \nonumber \\
&& \sum_{lm,l^{\prime}m^{\prime}} \int {{d^3k} \over {(2\pi)^3}}  
\exp\left[i{\bf k}\cdot ({\bf r}-{\bf r}_0) \right]
Y_{lm}(\hat{\bf s};\hat{\bf k}) 
\langle l m \vert K(k) \vert l^{\prime}m^{\prime} \rangle 
Y_{l^{\prime}m^{\prime}}^*(\hat{\bf s}_0;\hat{\bf k}) \ .
\label{G_Fourie_exp}
\end{eqnarray}

\noindent
Here $K(k)$ is an unknown operator. The reciprocity of the Green's
function, $G({\bf r},\hat{\bf s};{\bf r}_0,\hat{\bf s}_0) = G({\bf
  r}_0,-\hat{\bf s}_0;{\bf r},-\hat{\bf s})$, together with the fact
that $G$ is real, imply the following symmetry property of $K$:
$\langle l^{\prime} m^{\prime} \vert K \vert l m \rangle = (-1)^{l+
  l^{\prime}} \langle l m \vert K \vert l^{\prime} m^{\prime}
\rangle^*$. This can be also written as ${\mathcal
  P}K^{\dagger}{\mathcal P} = K$, where ${\mathcal P}$ is the
coordinate inversion operator with matrix elements $\langle lm \vert
{\mathcal P} \vert l^{\prime} m^{\prime} \rangle = (-1)^l\delta_{l
  l^{\prime}} \delta_{m m^{\prime}}$ and $\dagger$ denotes Hermitian
conjugation.  Thus, it can be seen that $K$ is not a Hermitian
operator.

Substituting (\ref{G_Fourie_exp}) into (\ref{RTE}) and using the
orthogonality properties of the spherical functions, we arrive at the
following operator equation for $K(k)$:

\begin{equation}
\label{K_matrix_eq}
(ikR + \Sigma)K(k) = 1 \ .
\end{equation}

\noindent
The matrices $R$ and $\Sigma$ are defined by

\begin{eqnarray}
\langle l m \vert R \vert l^{\prime}m^{\prime} \rangle && =
\int (\hat{\bf s} \cdot \hat{\bf k})
Y_{lm}^*(\hat{\bf s}; \hat{\bf k}) Y_{l^{\prime} m^{\prime}}(\hat{\bf
  s}; \hat{\bf k}) d^2\hat{\bf s} \nonumber \\
&& = \delta_{m m^{\prime}} \left[ b_{lm} \delta_{l^{\prime}=l-1} +
  b_{l+1,m} \delta_{l^{\prime}=l+1} \right] \ , 
\label{R_lmlm_def}\\
&& \hspace{-0.5cm} b_{lm} = \sqrt{ (l^2-m^2)/(4l^2-1)} \ ,
\label{b_lm_def}\\
\langle l m \vert \Sigma \vert l^{\prime}m^{\prime} \rangle && = \sigma_l
\delta_{l l^{\prime}} \delta_{m m^{\prime}} \ , 
\label{S_lmlm_def}
\end{eqnarray}

\noindent
where $\sigma_l$ is given by (\ref{sigma_l_def}). The formal solution
to (\ref{K_matrix_eq}) can be written as

\begin{equation}
\label{K_matrix_sol}
K(k) = S(1+ikW)^{-1}S \ ,
\end{equation}

\noindent
where $S=1/\sqrt{\Sigma}$ and $W=SRS$. Note that $\langle lm \vert S
\vert l^{\prime}m^{\prime} \rangle = \delta_{ll^{\prime}}
\delta_{mm^{\prime}} / \sqrt{\sigma_l}$ exists because $\sigma_l>0$,
which follows from the inequalities $A_l \leq 1$ and
$\mu_a>0$~\footnote[4]{Purely scattering media with $\mu_a=0$ can be
  considered separately.}.  Similarly to $R$, $W$ is a real symmetric
matrix. Therefore, we can use the spectral theorem to express
$(1+ikW)^{-1}$ in terms of the eigenvectors and eigenvalues of $W$,
$\vert \psi_\mu \rangle$ and $\lambda_\mu$, respectively. This
immediately leads to the following expression for $K(k)$:

\begin{equation}
\label{K_eig_sol}
K(k) = \sum_\mu {{S \vert \psi_\mu \rangle \langle
    \psi_\mu \vert S } \over {1 + ik\lambda_\mu}} \ ,
\end{equation}

\noindent
Given the set of eigenvectors and eigenvalues, which can be found by
numerical diagonalization of $W$, the above formula solves the problem
in Fourier space. Since the components of $\vert \psi_\mu \rangle$
in the $\vert lm \rangle$ basis are purely real, it can be seen that
$K$ is symmetric. 

\subsection{Mathematical properties of the solution}
\label{subsec:math_prop}

\subsubsection{Block structure of $W$\newline}
\label{subsubsec:block_structure}

First, we note that $W$ is block-diagonal: $\langle lm\vert W \vert
l^{\prime} m^{\prime}\rangle = \delta_{mm^{\prime}} \langle l \vert
B(m) \vert l^{\prime} \rangle$. Below, we will label different blocks
$B(M)$ ($M=0, \pm1, \pm2, \ldots$) by the capital letter $M$. The
matrix elements of $B(M)$ are given by

\begin{eqnarray}
\label{B_matrix_def}
&& \langle l \vert B(M) \vert l^{\prime} \rangle = 
\beta_l(M) \delta_{l^{\prime}=l-1} + \beta_{l+1}(M)
\delta_{l^{\prime}=l+1} \ 
, \ \ \ \ l,l^{\prime} \geq \vert M \vert \ , \\
\label{beta_l_def}
&& \beta_l(M) = b_{lM}/\sqrt{\sigma_l \sigma_{l-1}} \ .
\end{eqnarray}

\noindent
Obviously, to find all eigenvalues and eigenvectors of $W$, it is
sufficient to diagionalize each block separately. This task is further
simplified because all blocks $B(M)$ are tridiagonal. We denote
eigenvectors of a block $B(M)$ by $\vert \phi_n (M) \rangle$. Then the
eigenvector of the full matrix $W$ with the same eigenvalue is
obtained according to

\begin{equation}
\langle lm \vert \psi_{Mn} \rangle = \delta_{m
  M} \langle l \vert \phi_n(M) \rangle \ .
\end{equation}

\noindent
The corresponding eigenvalue is denoted by $\lambda_{Mn}$, where we
have introduced a composite index $(M,n)$. Note that $B(M)=B(-M)$.

\subsubsection{Symmetry properties of the eigenvectors\newline}
\label{subsubsec:symmetry}

The property ${\mathcal P}K^{\dagger}{\mathcal P} = K$ and
Eq.~(\ref{K_matrix_sol}) imply that ${\mathcal P}W{\mathcal P} = -W$.
Thus, $W$ is odd with respect to coordinate inversion. In particular,
if $\vert \psi \rangle$ is an eigenvector of $W$ with the eigenvalue
$\lambda$, then $\vert \tilde{\psi} \rangle = {\mathcal P} \vert \psi
\rangle$ is also an eigenvector of $W$ but with an eigenvalue of the
opposite sign, $\tilde{\lambda}=-\lambda$. The complete set of
eigenvectors $\{ \vert \psi_\mu \rangle\ : \mu \in \Omega \}$, where
$\Omega$ is the set of all values of the index $\mu$, can then be
equivalently rewritten as $\{ \vert \psi_\mu \rangle, \vert
\tilde{\psi}_\mu \rangle \ : \mu \in \Omega^{+} \}$, where $\Omega^+$
is the set of indices $\mu$ that correspond to positive eigenvalues
$\lambda_\mu$.  The set of indices that correspond to negative
eigenvalues can be denoted as $\Omega^-$; then $\Omega = \Omega^+ \cup
\Omega^-$ and $\Omega^+ \cap \Omega^- = \{0\}$.

Using these properties, one can transform the summation over all
values of $\mu$ in (\ref{K_eig_sol}) to summation over $\mu\in
\Omega^+$ (such sums will be denoted as ${\sum}_{\mu}^{\prime}$
below). This fact facilitates the inverse Fourier transformation
(see~\ref{App:A}) and solution of the boundary-value problem discussed
in Section~\ref{subsec:boundaries}.

\subsubsection{Continuous and discrete spectra\newline}
\label{subsubsec:cont_disc}

Third, the eigenvalues $\lambda_\mu$ can belong either to the discrete
or continuous spectrum. It is easy to see that the spectrum is
continuous for $\vert \lambda \vert < 1/\mu_t$, where $\mu_t= \mu_a +
\mu_s$, and discrete for $\vert \lambda \vert > 1/\mu_t$. Indeed,
consider the three-point recurrence relation that follows from the
equation $W \vert \psi \rangle = \lambda \vert \psi \rangle$:

\begin{equation}
\label{three-term}
\beta_l(m) \langle l-1,m \vert \psi \rangle  + \beta_{l+1}(m) \langle
l+1,m \vert \psi \rangle =  \lambda \langle l m \vert \psi \rangle \ ,
\ \ l\geq \vert m \vert \ .
\end{equation}

\noindent
In general, it has two types of solutions: polynomial and exponential.
Consider the asymptotic properties of these solutions.  In the limit
$l \rightarrow \infty$ we have $A_l\rightarrow 0$, $\sigma_l
\rightarrow \mu_t$, $b_{lm}\rightarrow 1/2$ and $\beta_l(m)
\rightarrow 1/2\mu_t$.  The recurrence relation then becomes:

\begin{equation}
\label{three-term_inf}
\langle l-1,m \vert \psi \rangle  + \langle l+1,m
\vert \psi \rangle =  2\mu_t \lambda \langle l m \vert \psi \rangle \ .
\end{equation}

\noindent
The polynomial solutions have the asymptotic form $\langle l m \vert
\psi \rangle = p_l^m(\lambda\mu_t)$, where $p_l^m(x)$ are general
orthogonal polynomials of degree $l$ (not to be confused with the
associated Legendre functions which solve the recurrence
(\ref{three-term}) in the particular case $\mu_s=0$).  In order for
this solution to be an eigenvector of $W$, it must be bounded.
Obviously, this requirement is equivalent to $\vert \lambda \mu_t
\vert \leq 1$. Thus, for every $\lambda \in [-1/\mu_t,1/\mu_t]$, there
is a polynomial solution to the three-term recurrence relation that is
an eigenvector of $W$.

For $\lambda$ outside of the interval $[-1/\mu_t,1/\mu_t]$, polynomial
solutions are unbounded and, therefore, can not be eigenvectors of
$W$.  We then consider exponential solutions which behave
asymptotically as $\langle lm\vert \psi \rangle = (\pm 1)^l \exp(-pl)$
where $p$ satisfies the equation $\cosh(p)=\pm \mu_t \lambda$. In
order for this solution to be an eigenvector of $W$, $p$ must be
positive. But the above equation has positive roots only when $\vert
\lambda \vert \geq 1/\mu_t$. Note that the exponentially decaying
eigenvectors have a finite $L^2$ norm, and, hence, belong to the
discrete spectrum. Further bounds on the discrete spectrum can be
inferred from the Gershgorin theorem, which states that, for a fixed
$M$, $\vert \lambda_{Mn} \vert \leq r_M = \max_{l \geq \vert M
  \vert}[\beta_l(M)+\beta_{l+1}(M)]$. It can be easily verified that
$r_0 = 4/\sqrt{3}\mu_a$ and $r_M=1/\mu_a$ for $\vert M \vert > 0$.

In numerical computations, the infinite-dimensional matrix $W$ must be
truncated and the continuous spectrum of $W$ approximated by a
discrete spectrum. In this paper we treat all eigenvalues as discrete.
Thus, for example, the expression (\ref{K_eig_sol}) contains only a
sum over discrete modes, although, theoretically, summation over the
continuous part of the spectrum must be expressed as an integral. Note
that an expression involving only discrete spectra avails itself more
readily to numerical implementation.

\subsection{Green's function in real space}
\label{subsec:Green_RS}

The dependence of solution (\ref{K_eig_sol}) on ${\bf k}$ is
analytical. This allows us to obtain the Green's function the
coordinate representation by Fourier transform. We substitute
(\ref{K_eig_sol}) into the ansatz (\ref{G_Fourie_exp}) and express the
spherical functions $Y_{lm}(\hat{\bf s};\hat{\bf k})$ and
$Y_{l^{\prime}m^{\prime}}(\hat{\bf s}_0;\hat{\bf k})$ in terms of
spherical functions defined in the laboratory frame whose $z$-axis
direction is given by a unit vector $\hat{\bf z}$ according to
(\ref{D_l}),(\ref{dmll}). The direction of the $x$- and $y$-axes of
the laboratory frame is arbitrary. This leads to the following
expression:

\begin{equation}
\label{G_chi_rs_gen}
G({\bf r}, \hat{\bf s}; {\bf r}_0, \hat{\bf s}_0) =
\sum_{lm} \sum_{l^{\prime} m^{\prime}}
Y_{lm}(\hat{\bf s}; \hat{\bf z}) \langle lm \vert \chi({\bf r}-{\bf r}_0;\hat{\bf z}) \vert l^{\prime}
m^{\prime} \rangle  
Y_{l^{\prime}m^{\prime}}^*(\hat{\bf s}_0; \hat{\bf z}) \ ,
\end{equation}

\noindent
where

\begin{equation}
\chi({\bf r};\hat{\bf z})  = 
\int {{d^3{\bf k} } \over {(2\pi)^3}} \exp(i{\bf k}\cdot {\bf r}) 
{\mathcal D}(\hat{\bf k};\hat{\bf z}) K(k) {\mathcal
  D}^{\dagger}(\hat{\bf k};\hat{\bf z})
\label{chi_rs_gen_oper}
\end{equation}

\noindent
Here ${\mathcal D}(\hat{\bf k};\hat{\bf z}) = \exp(-i\varphi_{\hat{\bf
    k}}J_z) \exp( -i\theta_{\hat{\bf k}}J_y)$ is the rotation operator
whose matrix elements are given by the Wigner D-functions, $\langle lm
\vert {\mathcal D}(\hat{\bf k};\hat{\bf z}) \vert l^{\prime}m^{\prime}
\rangle = \delta_{l l^{\prime}} D_{m m^{\prime}}^l (\varphi_{\hat{\bf
    k}}, \theta_{\hat{\bf k}}, 0)$, $\varphi_{\hat{\bf k}}$ and
$\theta_{\hat{\bf k}}$ are the polar angles of ${\bf k}$ in the
laboratory frame, and ${\bf J}$ is the angular momentum operator (with
$\hbar=1$).  We note that operators ${\mathcal D}$ are unitary and,
hence, normal: ${\mathcal D}^{-1}={\mathcal D}^{\dagger}$. However,
${\mathcal D}$ does not commute with $K$. The fundamental
simplification obtained by the MRRF is that ${\mathcal D}$ is known
analytically while $K$ has a simple form given by (\ref{K_eig_sol}).
In particular, given numerical values of $\vert \psi_\mu \rangle$ and
$\lambda_\mu$, the dependence of $K(k)$ on $k$ is also known
analytically.

Below, we consider two different cases. In the first case, the
direction of the laboratory frame $z$-axis coincides with the
direction from the source to the detector, namely, $\hat{\bf z}=({\bf
  r}-{\bf r}_0)/\vert{\bf r} - {\bf r}_0\vert$. This choice of the
angular basis is convenient when the source and the detector are
always placed on the same line, irrespective of the directions
$\hat{\bf s}$ and $\hat{\bf s}_0$. In the second case, we choose
$\hat{\bf z}=\hat{\bf s}_0$. This approach is useful when the source
is scanned, e.g., over a two-dimensional plane, but its direction
$\hat{\bf s}_0$ is fixed. This situation is typical for optical
tomography in the slab geometry. The integral (\ref{chi_rs_gen_oper})
for the two cases is evaluated in~\ref{App:A}. The result is, in the
first case:

\begin{eqnarray}
\langle lm \vert \chi({\bf r};\hat{\bf r}) \vert l^{\prime}m^{\prime}
\rangle = &&  { {\delta_{m m^{\prime}} } \over {2\pi \sqrt{\sigma_l \sigma_{l^{\prime}}} }}
\sum_{M=-\bar{l}}^{\bar{l}} (-1)^M \sum_{j=0}^{\bar{l}}
C_{l,M,l^{\prime},-M}^{\vert l - l^{\prime} \vert +2j,0} 
C_{l,m,l^{\prime},-m}^{\vert l - l^{\prime} \vert +2j,0} \nonumber \\
&& \times 
{\sum_\mu}^{\prime} {{\langle l M \vert \psi_\mu \rangle \langle \psi_\mu
    \vert l^{\prime} M \rangle} \over {\lambda_\mu^3}}
k_{\vert l - l^{\prime} \vert +2j}\left({R \over {\lambda_\mu}} \right) \ ,
\label{chi_llm_uni}
\end{eqnarray}

\noindent
Here $\bar{l} = \min (l,l^{\prime})$, $k_n(x)=-i^n h_n^{(1)}(ix)$ is
the modified spherical Bessel function of the first kind (defined
without a factor of $\pi/2$), $C_{j_1m_1j_2m_2}^{j_3m_3}$ are the
Clebsch-Gordan coefficients and ${\sum}^{\prime}$ denotes summation
over only such indices $\mu$ that correspond to positive eigenvalues
$\lambda_\mu$. It can be seen that $\chi({\bf r};\hat{\bf r})$ is
diagonal in $m$ and $m^{\prime}$, which corresponds to the invariance
of the Green's function with respect to a simultaneous rotation of the
vectors $\hat{\bf s}$ and $\hat{\bf s}_0$ around the line connecting
the source and the detector. Eq.~(\ref{chi_llm_uni}) can be further
simplified by expressing the eigenvectors $\vert \psi_\mu \rangle$ in
terms of the eigenvectors $\vert \phi_n(m) \rangle$ of the smaller
blocks $B(m)$ as discussed in Section~\ref{subsubsec:block_structure}.
This result, as well as a number of special cases, were given in
Ref.~\cite{markel_04_1} and are not repeated here.

In the case $\hat{\bf z}=\hat{\bf s}_0$, expression
(\ref{G_chi_rs_gen}) contains only matrix elements of $\chi({\bf
  r};\hat{\bf s}_0)$ with $m^{\prime}=0$. This follows from the fact
that $Y_{l^{\prime}m^{\prime}}(\hat{\bf s}_0; \hat{\bf
  s}_0)=\delta_{m^{\prime}0}\sqrt{(2l^{\prime}+1)/4\pi}$. The
corresponding expression for the matrix elements of $\chi({\bf
  r};\hat{\bf s}_0)$ is

\begin{eqnarray}
\label{chi_llm_uni_z}
\hspace{-2cm} \langle lm \vert \chi({\bf r};\hat{\bf s}_0) \vert l^{\prime} 0
\rangle &&  = { {(-1)^l} \over {\sqrt{\pi (2l^{\prime}+1) \sigma_l \sigma_{l^{\prime}}} }}
\sum_{M=-\bar{l}}^{\bar{l}} \sum_{j=0}^{\bar{l}}
Y_{\vert l - l^{\prime} \vert +2j,m}^*(\hat{\bf r};\hat{\bf s}_0) \nonumber \\
&& \times 
C_{l,M,\vert l - l^{\prime} \vert +2j,0}^{l^{\prime},M} 
C_{l,m,l^{\prime},0}^{\vert l - l^{\prime} \vert +2j,m} 
{\sum_\mu}^{\prime} {{\langle l M \vert \psi_\mu \rangle \langle \psi_\mu
    \vert l^{\prime} M \rangle} \over {\lambda_\mu^3}}
k_{\vert l - l^{\prime} \vert +2j} 
\left({R \over {\lambda_\mu}} \right) \ .
\end{eqnarray}

\noindent
Derivation of the above result is analogous to that for $\chi({\bf
  r};\hat{\bf r})$; see~\ref{App:A} for details.

\subsection{Plane-wave decomposition of the Green's function}
\label{subsec:pw_dec}

Having in mind further applications of the MRRF to solving boundary
value problems, we derive the plane-wave decomposition of the Green's
function. The latter is defined by the two-dimensional Fourier
integral

\begin{eqnarray}
G({\bf r},\hat{\bf s}; {\bf r}_0,\hat{\bf s}_0) = &&
\sum_{lm} \sum_{l^{\prime}m^{\prime}} \int {{d^2 q} \over { (2\pi)^2}} \exp[i{\bf
      q}\cdot({\bm \rho} - {\bm \rho}_0)] \nonumber \\
&& \times Y_{lm}(\hat{\bf s};\hat{\bf z}) 
\langle l m \vert \kappa({\bf q}; z-z_0) \vert l^{\prime} m^{\prime}
\rangle 
Y_{l^{\prime}m^{\prime}}^*(\hat{\bf s}_0;\hat{\bf z}) \ . 
\label{PW_dec_def}
\end{eqnarray}

\noindent
Here $\hat{\bf z}$ is a selected direction in space which coincides
with the $z$-axis of the laboratory frame, ${\bm \rho}$ is a
two-dimensional vector in the $x-y$ plane (${\bf r}={\bm
  \rho}+z\hat{\bf z}$ and ${\bm \rho} \cdot \hat{\bf z} = 0$) and the
direction of $x$- and $y$-axes is arbitrary.  By comparing the above
expression to (\ref{chi_rs_gen_oper}), we find that

\begin{equation}
\label{kappa_int_kz}
\hspace{-2cm} \kappa({\bf q};z) = \int_{-\infty}^{\infty} 
{{d k_z} \over {2\pi}} \exp(i k_z z)
{\mathcal D}({\bf q}+\hat{\bf z}k_z;\hat{\bf z}) K\left( \sqrt{q^2 +
    k_z^2} \right) {\mathcal D}^{\dagger}({\bf q}+\hat{\bf
  z}k_z;\hat{\bf z}) \ . 
\end{equation}

\noindent
Here ${\mathcal D}({\bf q}+\hat{\bf z}k_z;\hat{\bf z})$ should be
understood as a function of the polar angles of the vector ${\bf
  k}={\bf q} + \hat{\bf z}k_z$ in the laboratory frame. The latter are
defined by 

\begin{equation}
\label{cos_sin_theta}
\cos \theta = k_z/\sqrt{q^2 + k_z^2} \ , \ \ \ \sin \theta =
q/\sqrt{q^2 + k_z^2} \ .
\end{equation}

\noindent
Integral (\ref{kappa_int_kz}) can be evaluated analytically.  The
following expression for the matrix elements of $\kappa({\bf q};z)$ is
derived in~\ref{App:B}:

\begin{eqnarray}
\label{kappa_result}
\hspace{-2cm} \langle lm \vert \kappa({\bf q};z) \vert l^{\prime} m^{\prime} \rangle
= {{\exp[-i(m-m^{\prime})\varphi_{\hat{\bf q}}]} \over
  {\sqrt{\sigma_l\sigma_{l^{\prime}}}}} [{\rm sgn}(z)]^{l+l^{\prime}+m+m^{\prime}}
\sum_{m_1=-l}^l \sum_{m_2=-l^{\prime}}^{l^{\prime}}
{\sum_\mu}^{\prime} \nonumber \\
\hspace{-2cm} \times 
d_{m m_1}^l[i\tau(q\lambda_\mu)] 
\langle l m_1 \vert \psi_\mu \rangle
\frac{\exp\left[- Q_{\mu}(q) \vert z \vert \right]}
{\lambda_\mu^2 Q_{\mu}(q)}
\langle \psi_\mu \vert l^{\prime} m_2 \rangle 
d_{m^{\prime}m_2}^{l^{\prime}} [i\tau(q\lambda_\mu)] \ ,
\end{eqnarray}

\noindent
where 

\begin{equation}
\label{Q_def}
Q_{\mu}(q) = \sqrt{q^2 + 1/\lambda_{\mu}^2} \ ,
\end{equation}

\noindent
the complex angles $i\tau(x)$ are defined by the relations

\begin{equation}
\label{CS_tau_def}
\cos[i\tau(x)] = \sqrt{1 + x^2} \ , \ \ \sin[i\tau(x)] = -ix \ ,
\end{equation}

\noindent
and the angle $\varphi_{\hat{\bf q}}$ is the polar angle of the
two-dimensional vector ${\bf q}$ in the $x-y$ plane. The Wigner
d-functions $d_{m m^{\prime}}^l (i\tau)$ in the above expression are
algebraic functions of $\cos(i\tau)$ and $\sin(i\tau)$ (an explicit
expression in terms of Jacobi polynomials is given in~\ref{App:B}). An
expression for $ \kappa({\bf q};z)$ in terms of the block eigenvectors
$\vert \phi_n(M) \rangle$ which were introduced in
Section~\ref{subsubsec:block_structure} is also given in~\ref{App:B}.
Here we note some of the symmetry properties of the matrices
$\kappa({\bf q};z)$. Under inversion of the $z$-axis, we have
$\kappa({\bf q},-z)={\mathcal P}_z \kappa({\bf q},z){\mathcal P}_z$,
or, in component form,

\begin{equation}
\label{kappa_z_inversion}
\langle lm \vert \kappa({\bf q};-z) \vert l^{\prime} m^{\prime}
\rangle = (-1)^{l+l^{\prime}+m+m^{\prime}} \langle lm \vert
\kappa({\bf q};z) \vert l^{\prime} m^{\prime} \rangle \ .
\end{equation}

\noindent
Simultaneous inversion of the $x$- and $y$-axes (or, equivalently,
rotation around the $z$-axis by the angle $\pi$) is expressed as
$\kappa(-{\bf q},z)={\mathcal P}_{xy} \kappa({\bf q},z){\mathcal
  P}_{xy}$, or, in component form,

\begin{equation}
\label{kappa_xy_inversion}
\langle lm \vert \kappa(-{\bf q};z) \vert l^{\prime} m^{\prime}
\rangle = (-1)^{m+m^{\prime}} \langle lm \vert
\kappa({\bf q};z) \vert l^{\prime} m^{\prime} \rangle \ .
\end{equation}

We also note some particular cases of expressions (\ref{PW_dec_def})
and (\ref{kappa_result}). First, consider the case $\hat{\bf
  s}=\hat{\bf s}_0=\hat{\bf z}$. This corresponds to the source and
detector being oriented perpendicular to the surface of a slab. We
then use $Y_{lm}(\hat{\bf z};\hat{\bf
  z})=\delta_{m0}\sqrt{(2l+1)/4\pi}$ to obtain

\begin{eqnarray}
\hspace{-2.5cm}G({\bf r},\hat{\bf z}; {\bf r}_0,\hat{\bf z}) = &&
\sum_{l,l^{\prime}=0}^{\infty} {{\sqrt{(2l+1)(2 l^{\prime} + 1)}} \over
  {4\pi}}
\int {{d^2 q} \over { (2\pi)^2}} \exp[i{\bf q}\cdot({\bm \rho} - {\bm \rho}_0)]
\langle l 0 \vert \kappa({\bf q}; z-z_0) \vert l^{\prime} 0
\rangle \ . \nonumber \\
\label{PW_dec_def_s=s0=z}
\end{eqnarray}

\noindent
With the use of identity (\ref{Eq_A9a}) given below, $\langle l 0
\vert \kappa({\bf q}; z) \vert l^{\prime} 0 \rangle$ can be expressed
in terms of the associated Legendre functions of the first kind
$P_l^m(x)$ as

\begin{eqnarray}
\hspace{-2.5cm}\langle l 0 \vert \kappa({\bf q}; z)  && \vert l^{\prime} 0
\rangle =
{{[{\rm sgn}(z)]^{l+l^{\prime}}} \over
  {\sqrt{\sigma_l\sigma_{l^{\prime}}}}} \sum_{m_1=-l}^l
\sum_{m_2=-l^{\prime}}^{l^{\prime}} \sqrt{{(l-m_1)!(l^{\prime}-m_2)!} \over
  {(l+m_1)!(l^{\prime}+m_2)!}}{\sum_\mu}^{\prime} 
P_l^{m_1}\left[ \lambda_{\mu} Q_{\mu}(q) \right]
\nonumber \\ \hspace{-2cm} && \times 
\langle l m_1 \vert \psi_\mu \rangle
\frac{\exp\left[-Q_{\mu}(q) \vert z \vert \right]}{\lambda_\mu^2 Q_{\mu}(q)} 
\langle \psi_\mu \vert l^{\prime} m_2 \rangle 
P_{l^{\prime}}^{m_2}\left[ \lambda_{\mu} Q_{\mu}(q) \right] \ .
\end{eqnarray}

Next, consider the case ${\bf q}=0$. The operator $\kappa(0;z)$
describes one-dimensional propagation due to a planar source. We use
$d_{m m^{\prime}}^l(0)=\delta_{m m^{\prime}}$ to obtain

\begin{equation}
\label{kappa_result_q=0}
\hspace{-1cm} \langle l m \vert \kappa(0; z) \vert l^{\prime} m^{\prime} \rangle =
{{\delta_{m m^{\prime}} [{\rm sgn}(z)]^{l+l^{\prime}}} \over
  {\sqrt{\sigma_l\sigma_{l^{\prime}}}}} {\sum_\mu}^{\prime} \langle l
m \vert \psi_\mu \rangle \frac{\exp(-\vert z \vert /
  \lambda_\mu)}{\lambda_\mu} \langle \psi_\mu \vert l^{\prime} m^{\prime}
\rangle \ .
\end{equation}

\subsection{Plane wave and evanescent modes for the RTE}
\label{subsubsec:plane_eva}

Until now we considered solutions to the inhomogeneous RTE. However,
the solution of boundary value problems requires knowledge of the
general solution to the homogeneous equation. We seek such a solution
in the form $\exp(-{\bf k}\cdot{\bf r})\sum_{lm}\langle l m \vert c
\rangle Y_{lm}(\hat{\bf s};\hat{\bf k})$. Upon substitution of this
ansatz into the RTE with $\varepsilon=0$, we find that $\vert {\bf k}
\vert$ must be the generalized eigenvalue (with the direction of ${\bf
  k}$ being arbitrary) and $\vert c \rangle$ the generalized
eigenvector of the equation $k R\vert c \rangle = \Sigma\vert c
\rangle$, where the matrices $R$ and $\Sigma$ are defined by
(\ref{R_lmlm_def})-(\ref{S_lmlm_def}). Next we use (\ref{D_l}) to
express the spherical functions $Y_{lm}(\hat{\bf s};\hat{\bf k})$ in
terms of the spherical functions $Y_{lm}(\hat{\bf s};\hat{\bf z})$
defined in a laboratory frame with the $z$-axis pointing in a selected
direction. Then the general solutions to the homogeneous RTE
(\ref{RTE}) can be written as a superposition of the following modes:

\begin{equation}
\label{RTE_pw_mode}
\hspace{-2cm} I_{\hat{\bf k},M,n}({\bf r},\hat{\bf s})  = 
\exp \left(\frac{-\hat{\bf k}\cdot {\bf
  r}}{\lambda_{Mn}} \right) \sum_{lm} Y_{lm}(\hat{\bf s};\hat{\bf z})
\frac{\exp(-i m \varphi_{\hat{\bf k}})}{\sqrt{\sigma_l}}
d_{mM}^l(\theta_{\hat{\bf k}}) \langle l\vert \phi_n(M) \rangle \ .
\end{equation}

\noindent
Here it is more convenient to use the notation for block eigenvectors
$\vert \phi_n(M) \rangle$ which were introduced in
Section~\ref{subsubsec:block_structure}. The modes are labeled by the
unit vector $\hat{\bf k}$ ($\hat{\bf k} \cdot \hat{\bf k} =1$) whose
polar angles in the laboratory frame are $\theta_{\hat{\bf k}}$ and
$\varphi_{\hat{\bf k}}$ and by the composite index $\mu=(M,n)$. We
note that it is sufficient to consider only modes with positive
eigenvalues $\lambda_{Mn}$ ($\mu \in \Omega^+$; see
Section~\ref{subsubsec:symmetry}) due to the obvious symmetry
$I_{-\hat{\bf k},-M,n}(-{\bf r},\hat{\bf s}) = I_{\hat{\bf
    k},M,n}({\bf r},\hat{\bf s})$.

However, the modes (\ref{RTE_pw_mode}) with purely real vector
$\hat{\bf k}$ can not be used to construct a solution to a boundary
value problem in a half-space or in a slab. This is because each mode
is exponentially growing in the direction $-\hat{\bf k}$. Therefore,
it is necessary to define {\it evanescent} modes with complex-valued
vectors $\hat{\bf k}$. These modes are oscillatory in the $x-y$ plane
and exponentially decaying (or growing) in the $z$-direction. Namely,
let

\begin{equation}
\label{k_q_ev}
\hat{\bf k} = -i \lambda_{Mn}{\bf q} \pm \hat{\bf z} 
\lambda_{Mn}Q_{Mn}(q) \ ,
\end{equation}

\noindent
where ${\bf q} \cdot \hat{\bf z} = 0$. The polar angles of $\hat{\bf
  k}$ are defined as follows: $\varphi_{\hat{\bf k}} =
\varphi_{\hat{\bf q}}$, $\cos(\theta_{\hat{\bf k}}) = \hat{\bf k}
\cdot \hat{\bf z} = \pm \lambda_{Mn}Q_{Mn}(q)$ and
$\sin(\theta_{\hat{\bf k}}) = \hat{\bf k}\cdot \hat{\bf q} =
-iq\lambda_{Mn}$. Thus, we can write $\theta_{\hat{\bf k}} =
i\tau(q\lambda_{Mn})$, where the sine and cosine of the complex angle
$i\tau(x)$ are given by (\ref{CS_tau_def}). This gives rise to two
kinds of evanescent modes:

\begin{eqnarray}
\label{I_ev_plus}
\hspace{-2cm} I_{{\bf q},M,n}^{(+)}({\bf r},\hat{\bf s}) = && \exp\left[ i{\bf q}\cdot
  {\bm \rho} - Q_{\mu}(q) z \right]
\sum_{lm} Y_{lm}(\hat{\bf s};\hat{\bf z})
\frac{\exp(-i m \varphi_{\hat{\bf q}})} {\sqrt{\sigma_l}} \nonumber \\
&& \times d_{mM}^l[i\tau(q\lambda_{Mn})] \langle l \vert \phi_n(M) \rangle \ , \\ 
\label{I_ev_minus}
\hspace{-2cm} I_{{\bf q},M,n}^{(-)}({\bf r},\hat{\bf s}) = && (-1)^M\exp\left[ i{\bf q}\cdot
  {\bm \rho} + Q_{\mu}(q) z \right]
\sum_{lm} Y_{lm}(\hat{\bf s};\hat{\bf z})
\frac{\exp(-i m \varphi_{\hat{\bf q}})} {\sqrt{\sigma_l}} \nonumber \\
&& \times (-1)^{l+m}d_{m,-M}^l[i\tau(q\lambda_{Mn})] \langle l \vert \phi_n(M) \rangle \ .
\end{eqnarray}

\noindent
Here we have used the symmetry properties of $d_{mM}^l(\theta)$ under
the transformation $\theta \rightarrow \pi - \theta$ (which
corresponds to change of sign of the factor $\cos(\theta)$); hence the
additional phase factor $(-1)^{l+m}$ and the different sign of the
index $M$ in (\ref{I_ev_minus}). The phase factor $(-1)^M$ is
introduced for convenience.

The plane-wave decomposition (\ref{kappa_result}) can be equivalently
rewritten as an expansion in terms of evanescent waves:

\begin{equation}
\label{plane_wave_exp_evanescent}
G({\bf r},\hat{\bf s}; {\bf r}_0,\hat{\bf s}_0) =
{\sum_{\mu}}^{\prime} \int \frac{d^2 q}{(2\pi)^2} V_{{\bf q},\mu} \ 
I_{{\bf q},\mu}^{(\pm)} ({\bf r},\hat{\bf s}) \ 
I_{-{\bf q},\mu}^{(\mp)} ({\bf r}_0,-\hat{\bf s}_0) \ ,
\end{equation}

\noindent
where

\begin{equation}
\label{V_def}
V_{{\bf q},\mu} = \frac{1}{\lambda_{\mu}^2 Q_{\mu}(q)}
\end{equation}

\noindent
and the upper signs must be selected if $z > z_0$ while the lower
signs are selected if $z< z_0$. It can be easily verified that the
expression (\ref{plane_wave_exp_evanescent}) obeys the reciprocity
condition.

Now we make the following observation. Evanescent waves propagating in
different directions can not, in principle, have identical angular
dependence. In particular, by analyzing only the angular dependence of
specific intensity in the plane $z=0$ (in infinite space), it is
possible to tell if this intensity was produced by sources located to
the left of the observation plane (in the region $z<0$), or to the
right. To demonstrate this point, we introduce vectors $\vert
\eta_{\mu}(q) \rangle = \vert \eta_{Mn}(q) \rangle$ with components
$\langle l m \vert \eta_{Mn}(q) \rangle =
d_{mM}^l[i\tau(q\lambda_{Mn})]\langle l \vert \phi_n(M) \rangle$ and a
set of vectors obtained from $\vert \eta_{\mu}(q) \rangle$ by
coordinate inversion: $\vert \tilde \eta_{\mu}(q) \rangle = {\mathcal
  P} \vert \eta_{\mu}(q) \rangle$. The evanescent modes
(\ref{I_ev_plus}),(\ref{I_ev_minus}) can be written as

\begin{eqnarray}
\label{I_ev_plus_eta}
\hspace{-2cm} I_{{\bf q},\mu}^{(+)}({\bf r},\hat{\bf s}) = && \exp\left[ i{\bf q}\cdot
  {\bm \rho} - Q_{\mu}(q) z \right]
\sum_{lm} Y_{lm}(\hat{\bf s};\hat{\bf z}) \langle l m \vert {\mathcal
  A}(\hat{\bf q}) \vert \eta_{\mu}(q) \rangle \ , \\ 
\label{I_ev_minus_eta}
\hspace{-2cm} I_{{\bf q},\mu}^{(-)}({\bf r},\hat{\bf s}) = &&
\exp\left[ i{\bf q}\cdot {\bm \rho} + Q_{\mu}(q) z \right]
\sum_{lm} Y_{lm}^*(\hat{\bf s};\hat{\bf z}) \langle l m \vert {\mathcal
  A}^{\dagger}(-\hat{\bf q}) \vert \tilde{\eta}_{\mu}(q) \rangle \ , \\ 
&& \langle l m \vert {\mathcal A}(\hat{\bf q}) \vert l^{\prime}
m^{\prime} \rangle = \delta_{l l^{\prime}} \delta_{m m^{\prime}}
\exp(-im \varphi_{\hat{\bf q}})/\sqrt{\sigma_l}
\end{eqnarray}

\noindent
Here ${\mathcal A}(\hat{\bf q})$ is a diagonal matrix. Note that
${\mathcal A}(\hat{\bf q})$ depends only on the direction of the real
vector ${\bf q}$, while $\vert \eta_{\mu}(q) \rangle$ depends only on
its length.  In the case $q=0$ we have $\vert \eta_{\mu}(0) \rangle =
\vert \psi_{\mu} \rangle$ and $\vert \tilde{\eta}_{\mu}(0) \rangle =
\vert \tilde{\psi}_{\mu} \rangle$. Thus, the set $\{ \vert
\eta_{\mu}(0) \rangle,\ \vert \tilde{\eta}_{\mu}(0) \rangle \}$ forms
an orthonormal basis in the Hilbert space spanned by the eigenvectors
of $W$, ${\mathcal H}$. In the case $q\neq 0$, the vectors $\{ \vert
\eta_{\mu}(q) \rangle,\ \vert \tilde{\eta}_{\mu}(q) \rangle \}$ are no
longer orthonormal. However, we believe on physical grounds that they
still form a basis in ${\mathcal H}$~\footnote[5]{We do not have a
  direct proof of this statement. However, in the opposite case, the
  boundary value problem would not be uniquely solvable.}. Then there
exists a {\it dual basis} $\{ \vert \zeta_{\mu}(q) \rangle , \ \vert
\tilde{\zeta}_{\mu}(q) \rangle \}$ such that $\langle \zeta_{\mu}(q)
\vert \eta_{\nu}(q) \rangle = \delta_{\mu \nu}$, $\langle
\tilde{\zeta}_{\mu}(q) \vert \tilde{\eta}_{\nu}(q) \rangle =
\delta_{\mu \nu}$ and $\langle \tilde{\zeta}_{\mu}(q) \vert
\eta_{\nu}(q) \rangle = \langle \zeta_{\mu}(q) \vert
\tilde{\eta}_{\nu}(q) \rangle = 0$.

Now assume that we have measured the specific intensity in the plane
$z=0$. Denote the two-dimensional Fourier transform of this function
with respect to $x$ and $y$ by $I_0({\bf q},\hat{\bf
  s})=\sum_{lm}I_{lm}({\bf q}) Y_{lm}(\hat{\bf s};\hat{\bf z})$. The
expansion coefficients $I_{lm}({\bf q})$ are elements of a vector
$\vert I({\bf q}) \rangle$. For every value of ${\bf q}$, we can form
a vector ${\mathcal A}^{-1}(\hat{\bf q}) \vert I({\bf q}) \rangle$,
since ${\mathcal A}(\hat{\bf q})$ is invertible. If the sources are
located to the left of the measurement plane, then $\langle
\tilde{\zeta}_{\mu}(q) \vert A^{-1}(\hat{\bf q}) \vert I({\bf q})
\rangle = 0$ for every $\mu\in \Omega^+$. In other words, the
projection of ${\mathcal A}^{-1}(\hat{\bf q}) \vert I({\bf q})
\rangle$ onto the dual subspace spanned by $\vert \zeta_{\mu}(q)
\rangle$ is equal to zero.  Analogously, if the sources are located to
the right of the observation plane, $\langle \zeta_{\mu}(q) \vert
A^{-1}(\hat{\bf q}) \vert I({\bf q}) \rangle = 0$. Since the
projection of a nonzero vector onto both subspaces can not be
simultaneously zero, the angular dependence of the specific intensity
in the plane $z=0$ carries information about the location of the
source with respect to this plane. We emphasize that this analysis is
only valid in the absence of boundaries.

\subsection{Boundary value problem}
\label{subsec:boundaries}

Any solution to the RTE in a half-space or in a slab can be
constructed as an expansion over modes
(\ref{I_ev_plus_eta}),(\ref{I_ev_minus_eta}) with indices $\mu$
corresponding to only positive eigenvalues $\lambda_{\mu}$. This fact
is crucial for application of the MRRF to the boundary value problem,
which in radiative transport theory is formulated in the {\it
  half-range} of the angular variable. Now we demonstrate how it can
be used to construct a solution to the boundary value problem posed on
one or two planar interfaces.

\subsubsection{External source incident on a half-space\newline}
\label{subsubsec:ext_source_z=0}

Consider the RTE in the half-space $z>0$. In this Section we assume
that there are no {\it internal sources} in the medium, i.e., the RTE
has a zero source term. The presence of external sources is expressed
through an inhomogeneous boundary condition at the interface $z=0$:

\begin{equation}
\label{bc_z=0}
I_0({\bm \rho},\hat{\bf s}) = I_{\rm inc}({\bm \rho},\hat{\bf s}) \ ,
\ \ {\rm if} \ \hat{\bf s}\cdot \hat{\bf z} > 0 \ .
\end{equation}

\noindent
Here $I_0({\bm \rho},\hat{\bf s})$ is the specific intensity evaluated
at $z=0$ and $I_{\rm inc}({\bm \rho},\hat{\bf s})$ is the intensity
incident from vacuum (the external source). The boundary condition
(\ref{bc_z=0}) is formulated in the half-range of the singular
variable.

The general solution to the RTE in the half-space $z>0$ can be written
as a superposition of outgoing evanescent waves of the form
(\ref{I_ev_plus_eta}):

\begin{equation}
\label{I_I_ev_z>0}
I({\bf r}, \hat{\bf s}) = \int \frac{d^2q}{(2\pi)^2}
{\sum}^{\prime}_{\mu} F_{{\bf q},\mu}^{(+)} I_{{\bf q},\mu}^{(+)}({\bf
  r},\hat{\bf s}) \ ,
\end{equation}

\noindent
where the unknown coefficients $F_{{\bf q},\mu}^{(+)}$ must be found
from the boundary condition (\ref{bc_z=0}). Now we use expansion
(\ref{I_I_ev_z>0}) and expression (\ref{I_ev_plus_eta}) to calculate
$I_0({\bm \rho},\hat{\bf s})$. Upon Fourier transformation of
(\ref{bc_z=0}) with respect to ${\bm \rho}$, we arrive at the
following equation:

\begin{equation}
\label{bc_1}
\sum_{lm} {\sum}^{\prime}_{\mu} Y_{lm}(\hat{\bf s};\hat{\bf z}) \langle l m
\vert {\mathcal A}(\hat{\bf q}) \vert \eta_{\mu}(q) \rangle F_{{\bf
    q},\mu}^{(+)} = I_{\rm inc}({\bf q},\hat{\bf s}) , \ \ {\rm if} \
\hat{\bf s}\cdot \hat{\bf z} > 0 .
\end{equation}

\noindent
Next, we multiply both sides of Eq.~\ref{bc_1} by
$Y_{l^{\prime}m^{\prime}}^*(\hat{\bf s}; \hat{\bf z})$ and integrate
over all directions such that $\hat{\bf s} \cdot \hat{\bf z} > 0$.
Note that integration in the right-hand side can be extended to all
directions of $\hat{\bf s}$ since $I_{\rm inc}({\bf q},\hat{\bf s})$
is identically zero for $\hat{\bf s} \cdot \hat{\bf z} < 0$. Thus, for
a collimated narrow incident beam which crosses the boundary at ${\bm
  \rho} = {\bm \rho}_0$ in the direction $\hat{\bf s}_0$, we obtain

\begin{equation}
\label{bc_2}
{\sum}^{\prime}_{\mu} \langle l m \vert {\mathcal B}{\mathcal
  A}(\hat{\bf q}) \vert \eta_{\mu}(q) \rangle F_{{\bf q},\mu}^{(+)} =
\exp(-i{\bf q}\cdot {\bm \rho}_0) Y_{lm}^*(\hat{\bf s}_0;\hat{\bf z})
\ ,
\end{equation}

\noindent
where matrix ${\mathcal B}$ is given by

\begin{eqnarray}
\hspace{-2cm} \langle l m \vert {\mathcal B} \vert l^{\prime} m^{\prime} \rangle && =
\int_{\hat{\bf s} \cdot \hat{\bf z} > 0} Y_{lm}^*(\hat{\bf s};\hat{\bf
  z}) Y_{l^{\prime}m^{\prime}}(\hat{\bf s};\hat{\bf
  z}) d^2 s \nonumber \\
&& = \frac{\delta_{m m^{\prime}}}{2}
\sqrt{\frac{(2l+1)(2l^{\prime}+1)(l-m)!(l^{\prime}-m)!}{(l+m)!(l^{\prime}+m)!}}
\int_0^1 P_l^m(x) P_{l^{\prime}}^m (x) dx \ .
\label{B_def}
\end{eqnarray}

For a fixed value of ${\bf q}$, (\ref{bc_2}) is a set of linear
equations of infinite size. In practice, this set must be truncated so
that $l<l_{\rm max}$. Then the number of equations is $2N=(l_{\rm
  max}+1)^2$, where we have assumed for simplicity that $l_{\rm max}$
is odd. But the number of unknowns $F_{{\bf q},\mu}^{(+)}$ is only
equal to $N$, since $\mu \in \Omega^+$. Therefore, (\ref{bc_2}) is
formally overdetermined. However, not all equations in (\ref{bc_2})
are linearly independent. In fact, the rank of ${\mathcal B}$ is
exactly equal to half of its size, which is a consequence of
half-range integration in (\ref{B_def}). Therefore we come to the
conclusion that (\ref{bc_2}) is a well-determined system of equations
with respect to the $N$ unknowns $F_{{\bf q},\mu}^{(+)}$.

Numerically, the problem of solving (\ref{bc_2}) can be solved in two
different ways. A direct approach is to consider only equations in
(\ref{bc_2}) which are linearly independent. This is achieved by only
leaving equations in the system with $l$ having the same parity as
$m$, e.g., for a fixed $m$, $l=\vert m \vert,\vert m \vert + 2, \vert
m \vert + 4, \ldots$, with the restriction $l \leq l_{\rm max}$.
Another approach is to seek the generalized Moore-Penrose
pseudoinverse of (\ref{bc_2}). In this case eigenvectors of the
truncated matrix ${\mathcal B}$ must be found numerically. If the size
of ${\mathcal B}$ is even, half of its eigenvalues will be zero. Let
$\vert \xi_{\nu} \rangle$ be the eigenvectors of ${\mathcal B}$ with
nonzero eigenvalues $\alpha_{\nu}$. Then the system of equations
(\ref{bc_2}) can be rewritten as

\begin{equation}
\label{bc_3}
\hspace{-.5cm} \alpha_{\nu} {\sum}^{\prime}_{\mu}  \langle \xi_{\nu} \vert {\mathcal A}(\hat{\bf q}) 
\vert \eta_{\mu}(q) \rangle F_{{\bf q},\mu}^{(+)} =
\exp(-i{\bf q}\cdot {\bm \rho}_0) \sum_{lm} \langle \xi_{\nu} \vert l
m \rangle Y_{lm}^*(\hat{\bf s}_0;\hat{\bf z}) \ .
\end{equation}

\noindent
We note that in the limit $l_{\rm max}\rightarrow \infty$, the
eigenvectors of ${\mathcal B}$ are known and are of simple form:
$\langle l m \vert \xi_{\hat{\bf s}} \rangle = Y_{lm}^*(\hat{\bf s};
\hat{\bf z})$ with the eigenvalues being unity for $\hat{\bf s}\cdot
\hat{\bf z} > 0$ and zero otherwise, i.e., ${\mathcal B}$ is
idempotent.

The system (\ref{bc_3}) can be simplified by the substitution $F_{{\bf
    q},\mu}^{(+)} = f_{{\bf q},\mu}^{(+)} \exp(-i{\bf q}\cdot {\bm
  \rho}_0)$. The coefficients $f_{{\bf q},\mu}^{(+)}$ are then
independent of the source coordinate ${\bm \rho}_0$. Another
simplification is achieved by noting that both ${\mathcal A}$ and
${\mathcal B}$ are diagonal in indices $m$ and $m^{\prime}$.  In
effect, the system (\ref{bc_3}) must be solved once for each value of
$\vert {\bf q} \vert$; the dependence of the solution on the direction
of ${\bf q}$ is trivial. If, in addition, the incident beam is normal
to the interface ($\hat{\bf s}_0=\hat{\bf z}$), the solutions do not
depend on $\hat{\bf q}$ at all.

The additional computational complexity associated with solving the
boundary value problem is then as follows. For every numerical value
of the lengths of the vector ${\bf q}$ which is used in the expansion
(\ref{I_I_ev_z>0}), a system of linear equations of size $N=(l_{\rm
  max}+1)^2$ must be solved (the cost of diagonalization of ${\mathcal
  B}$ is negligibly small).  Thus, consideration of boundary
conditions adds significant computational complexity to the problem.
This is a consequence of the fact that the rotation matrices
$\exp[\tau(q\lambda_{\mu})J_y]$, unlike the matrix $W$, are not
diagonal in $m$ and $m^{\prime}$. As a result, the system of equations
(\ref{bc_3}) is not block diagonal and, in addition, $q$-dependent.
However, the problem is easily solvable for $l_{\rm max} \leq 100$,
which is, perhaps, more than is needed in any practical computation.

\subsubsection{External source incident on a slab\newline}
\label{subsubsec:ext_source_z=0_Z=L}

The generalization of the mathematical apparatus developed in
Section~\ref{subsubsec:ext_source_z=0} to the case of the RTE in a
finite slab is straightforward. Consider RTE in the slab $0<z<L$. The
external source is assumed to be incident from the left. Then the
boundary conditions read

\begin{eqnarray}
\label{bc_Z=0_z=L}
&& I_0({\bm \rho},\hat{\bf s}) = I_{\rm inc}({\bm \rho},\hat{\bf s}) \ ,
\ \ {\rm if} \ \hat{\bf s}\cdot \hat{\bf z} > 0 \ , \\
\label{bc_z=0_Z=L}
&& I_L({\bm \rho},\hat{\bf s}) = 0 \ ,
\ \ {\rm if} \ \hat{\bf s}\cdot \hat{\bf z} < 0 \ ,
\end{eqnarray}

\noindent
where $I_0$ and $I_L$ are the specific intensities evaluated at the
surfaces $z=0$ and $z=L$, respectively. The general solution inside
the slab has the form

\begin{equation}
\label{I_I_ev_0<z<L}
I({\bf r}, \hat{\bf s}) = \int \frac{d^2q}{(2\pi)^2}
{\sum}^{\prime}_{\mu} \left[ 
F_{{\bf q},\mu}^{(+)} I_{{\bf q},\mu}^{(+)}({\bf r},\hat{\bf s}) +
F_{-{\bf q},\mu}^{(-)} I_{-{\bf q},\mu}^{(-)}({\bf r},-\hat{\bf s}) 
\right] \ ,
\end{equation}

\noindent
where $F_{{\bf q},\mu}^{(+)}$ and $F_{{\bf q},\mu}^{(-)}$ are unknown
coefficients. After some manipulations, we arrive at the following
system of equations:

\begin{eqnarray}
\hspace{-2cm}&& {\sum}^{\prime}_{\mu} \left\{
\langle l m \vert {\mathcal B}{\mathcal  A}(\hat{\bf q}) \vert \eta_{\mu}(q) 
\rangle F_{{\bf q},\mu}^{(+)} +
\exp\left[-Q_{\mu}(q)L\right]\langle l,-m \vert {\mathcal B}{\mathcal A}^{\dagger}(\hat{\bf q}) \vert \eta_{\mu}(q) 
\rangle F_{{\bf q},\mu}^{(-)} \right\} = \nonumber \\
\hspace{-2cm}&&\hspace{6cm}\exp(-i{\bf q}\cdot {\bm \rho}_0) Y_{lm}^*(\hat{\bf s}_0;\hat{\bf z}) \ , 
\label{bc_2a} \\
\label{bc_2b}
\hspace{-2cm}&& {\sum}^{\prime}_{\mu} \left\{ \exp\left[-Q_{\mu}(q)L\right]
\langle l m \vert {\mathcal B}{\mathcal  A}(\hat{\bf q}) \vert \tilde{\eta}_{\mu}(q) \rangle 
F_{{\bf q},\mu}^{(+)} + 
\langle l,-m \vert {\mathcal B}{\mathcal A}^{\dagger}(\hat{\bf q}) \vert \tilde{\eta}_{\mu}(q)
\rangle F_{{\bf q},\mu}^{(-)} \right\} = 0 \ .
\end{eqnarray}

\noindent
This set of equations is the analog of (\ref{bc_2}) for the case of a
finite slab. In the limit $L\rightarrow\infty$ one has $F_{{\bf
    q},\mu}^{(-)}=0$ and (\ref{bc_2a}) coincides with (\ref{bc_2}). We
note that for a fixed ${\bf q}$, (\ref{bc_2a}),(\ref{bc_2b}) is a set
of $2N$ linearly independent equations for $2N$ unknowns. The methods
briefly discussed in Section~\ref{subsubsec:ext_source_z=0} can be
used to obtain the solution.

\subsubsection{Internal source in a half-space\newline}
\label{subsubsec:int_source}

Next, we consider an internal source in the half-space $z>0$. Here we
assume that there are no external sources. However, if this is not so,
the solution can be obtained by simple superposition. 

Consider a point unidirectional source of the form $\varepsilon =
\delta({\bm \rho} - {\bm \rho}_0)\delta(z-z_0)\delta(\hat{\bf s}-
\hat{\bf s}_0)$, where $z_0>0$. The general solution in the region
$0<z<z_0$ is written as 

\begin{equation}
\label{I_half_space_int_source}
\hspace{-1cm} I({\bf r},\hat{\bf s}) =
\int \frac{d^2 q}{(2\pi)^2} {\sum_{\mu}}^{\prime} 
\left[
V_{{\bf q},\mu} 
I_{ {\bf q},\mu}^{(-)} ({\bf r},   \hat{\bf s}) 
I_{-{\bf q},\mu}^{(+)} ({\bf r}_0,-\hat{\bf s}_0) + 
F_{{\bf q},\mu}^{(+)} 
I_{{\bf q},\mu}^{(+)} ({\bf r},\hat{\bf s}) 
\right] \ ,
\end{equation}

\noindent
The second term in the square brackets in the right-hand side of the
above expression can be interpreted as the surface term in the
Kirchhoff-type formula for the Green's function (for the formulation of
the Kirchhoff integral specific to the RTE see~\cite{case_67_book}
or, for a more detailed derivation,~\cite{case_69_proceedings}). The
boundary condition at the interface $z=0$ reads:

\begin{equation}
\label{bc_z=0_homo}
I_0({\bm \rho},\hat{\bf s}) = 0 \ ,
\ \ {\rm if} \ \hat{\bf s}\cdot \hat{\bf z} > 0 \ .
\end{equation}

\noindent
The fact that the boundary condition is homogeneous reflects the fact
that there are no external sources. The latter can be included by
considering an inhomogeneous boundary condition of the type
(\ref{bc_z=0}). By analogy with
Section~\ref{subsubsec:ext_source_z=0}, we immediately arrive at the
following set of equations for the unknown coefficients $F_{{\bf
    q},\mu}^{(+)}$:

\begin{equation}
\label{bc_2c}
\hspace{-2cm} {\sum}^{\prime}_{\mu} \left[  
\langle l m \vert {\mathcal B}{\mathcal A}(\hat{\bf q}) \vert \eta_{\mu}(q) 
\rangle F_{{\bf q},\mu}^{(+)} 
+ V_{{\bf q},\mu} 
\langle l,-m \vert {\mathcal B}{\mathcal A}^{\dagger}(\hat{\bf q})
\vert \eta_{\mu}(q) \rangle
I_{-{\bf q},\mu}^{(+)} ({\bf r}_0,-\hat{\bf s}_0) 
\right] = 0 \ .
\end{equation}

\noindent
Similarly to (\ref{bc_2}), this is a set of $N$ linearly independent
equations with respect to $N$ unknowns.

\section{Numerics}
\label{sec:num}

Now we illustrate the expressions obtained in
Section~\ref{subsec:Green_RS} for the RTE Green's function in an
infinite medium with several numerical examples. We have computed the
Green's function by truncating the series in (\ref{G_chi_rs_gen}) at
$l,l^{\prime} \leq l_{\rm max}$ and using the reference frame in which
the $z$-axis is aligned with the direction of the source, $\hat{\bf
  s}_0$. Correspondingly, the expression (\ref{chi_llm_uni_z}) was
used to compute the matrix elements of $\chi$.  Note that in this
expression the summation over $M$ and $j$ is finite; however, the
summation over the modes $\vert \psi_\mu \rangle$ is infinite and must
be truncated. We have found empirically that, for each block $B(M)$ of
the matrix $W$, summation over $N=500$ eigenmodes (which corresponds
to $1,000\times 1,000$ matrices $B(M)$) is sufficient for all cases
shown below.  Further increase of $N$ does not change the result
within double precision machine accuracy.  The results start to
deviate noticeably from those computed at $N=500$ when $N$ is taken to
be smaller than $\sim 100$, especially when $l_{\rm max}$ is
relatively large.  Further, we have used the Henyey-Greenstein model
for the phase function, so that $A_l = g^l$ were $0<g<1$ is a
parameter.  In all figures shown below we calculate the specific
intensity $I({\bf r},\hat{\bf s})$ due to a point unidirectional
source placed at the origin and illuminating in the $z$-direction. The
distance from the source is measured in units of the transport free
path, $\ell^* = 1/[\mu_a + (1-g)\mu_s]$, which plays an important role
in diffusion theory.

It should be noted that numerical implementation of the formulas
derived in this paper requires a degree of caution because the Green's
function of the RTE is not square integrable with respect to both of
its arguments, ${\bf r}$ and $\hat{\bf s}$. Therefore, one can not
expect uniform point-wise convergence of the result with $l_{\rm
  max}$.  Mathematically, this is manifested in the fact that the
Bessel functions $k_l(x)$ that enter into
(\ref{chi_llm_uni}),(\ref{chi_llm_uni_z}) diverge factorially for
large orders: $k_l(x) \propto l!! \ (l\rightarrow \infty)$. This
growth can not be compensated either by the Clebsch-Gordan coefficients,
or by the eigenvector components $\langle lm \vert \psi_\mu \rangle$
which decay, at best, exponentially (see discussion in
Section~\ref{subsubsec:cont_disc}). Therefore,
(\ref{G_chi_rs_gen}),(\ref{chi_llm_uni}),(\ref{chi_llm_uni_z}) must be
viewed as expressions defining the {\em moments} of the Green's
function and the latter as a generalized function or a distribution.
Nevertheless, in most practical situations, the spatial and angular
dependencies of the Green's function can be approximated by smooth
square-integrable functions by truncating (\ref{G_chi_rs_gen}) at
certain values of $l_{\rm max}$ that provide desirable angular
resolution. Computations are further facilitated by analytical
subtraction of the {\em ballistic component} of the Green's function:

\begin{equation}
\label{ball_0}
G_b({\bf r}, \hat{\bf s}; {\bf r}_0,\hat{\bf s}_0) =
\delta{(\hat{\bf s} - \hat{\bf s}_0)}\delta{(\hat{\bf R} - \hat{\bf
    s}_0)} {{\exp(-\mu_t R)} \over {R^2}} \ ,  \ \ {\bf R}={\bf
  r}-{\bf r}_0 \ .
\end{equation}

\noindent
The corresponding ballistic contribution to $\chi_b$ is

\begin{equation}
\label{B1}
\langle l m \vert \chi_b(R;\hat{\bf s}_0) \vert l^{\prime}
m^{\prime} \rangle  = \delta_{m0} \sqrt{(2l +
  1)(2l^{\prime} + 1)} { {\exp(-\mu_t  R)} \over {4\pi R^2} }  \ , 
\end{equation}

\noindent
However, it is impossible to remove the singularities {\em
  completely}, and the remainder of such a subtraction still remains
non-square integrable.

\begin{figure}
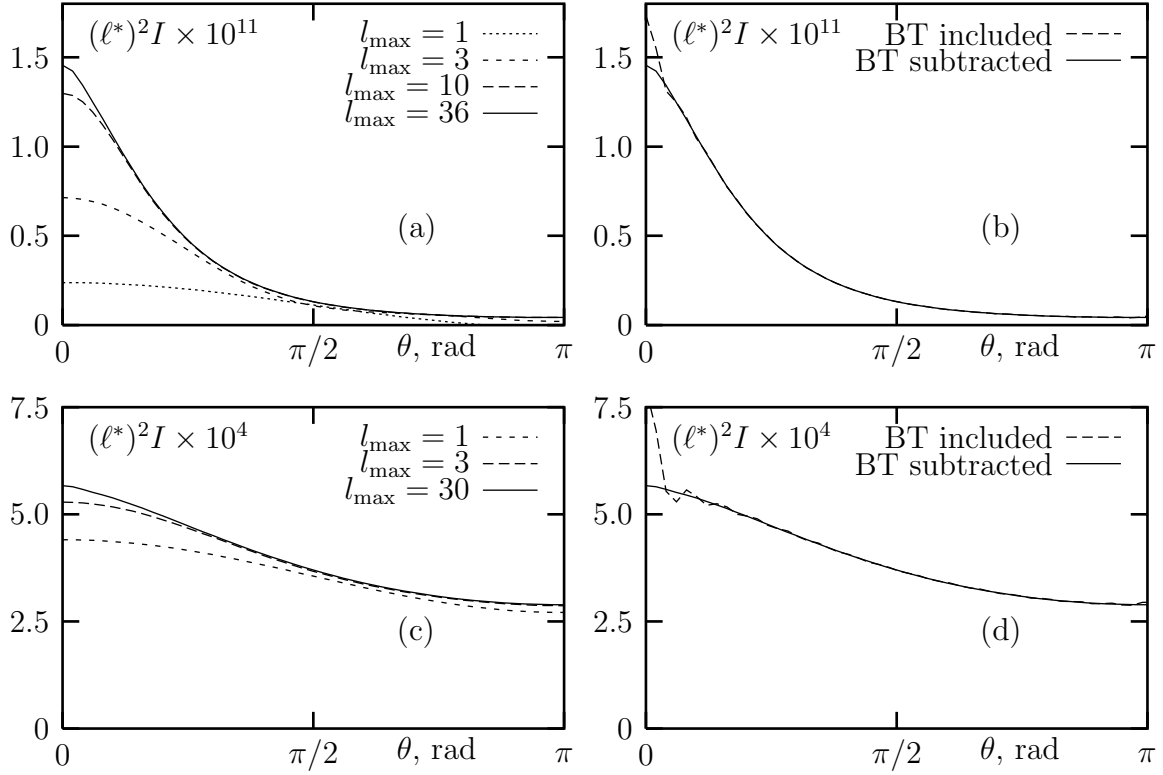

\centerline{\input{fig100a.tex}\hspace{-0.5cm}\input{fig100b.tex}}
\centerline{\input{fig101a.tex}\hspace{-0.5cm}\input{fig101b.tex}}
\caption{\label{fig:bt_subtr}Angular dependence of the specific
  intensity for forward propagation at the distance $z$ from the
  source. Left column (a,c): convergence with parameter $l_{\rm max}$.
  Right column (b,d): solid line shows the converged result obtained
  with subtraction of the ballistic term; dashed line: result obtained
  without subtraction of the ballistic term for the same $l_{\rm
    max}$. Top row (a,b): $g=0.5$, $\mu_a/\mu_s=0.5$, $z=20\ell^*$.
  Bottom row (c,d): $g=0.2$, $\mu_a/\mu_s=0.01$ and $z=10\ell^*$.}
\end{figure}

The effect of subtraction of the ballistic term and convergence with
$l_{\rm max}$ for forward propagation is illustrated in
Fig.~\ref{fig:bt_subtr}. Here $\theta$ is the angle between the
direction of observation, $\hat{\bf s}$, and the positive direction of
the $z$-axis: $\cos\theta=\hat{\bf s}\cdot \hat{\bf z}$. In the left
column (plots a,c) we show the dependence of the specific intensity
(with the ballistic term subtracted) on the maximum order of spherical
functions $l_{\rm max}$. We assumed that convergence was reached when
incrementing $l_{\rm max}$ by $1$ resulted in less than $0.1\%$
relative change of the specific intensity in any direction. However,
we emphasize again that this convergence is asymptotic. In the right
column of images (b,d), we compare the angular dependence of the
specific intensity for the maximum value of $l_{\rm max}$ which was
used in the graph to the left with and without the ballistic term.
Note that the subtracted ballistic term can be added back analytically
to the solutions obtained. In all figures shown below, the ballistic
term is subtracted.

\begin{figure}
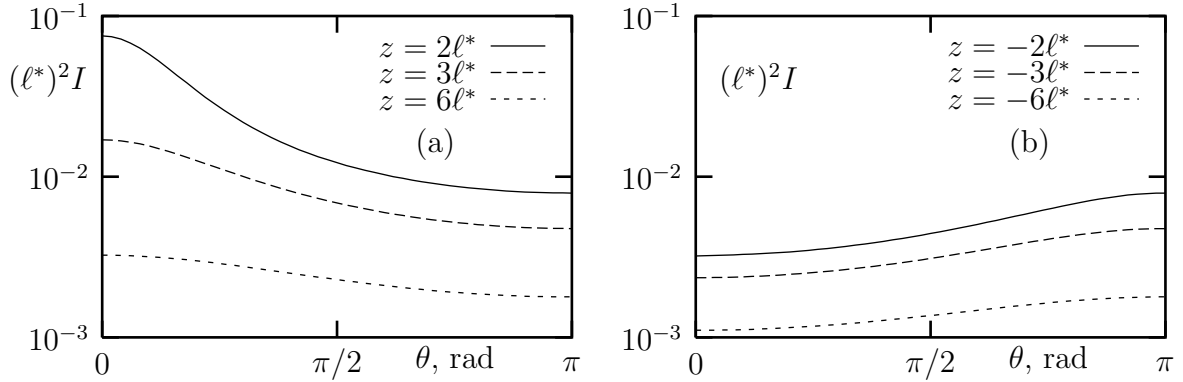

\centerline{\input{figa_onaxis.tex}\hspace{-1.0cm}\input{figb_onaxis.tex}}
\caption{\label{fig:onaxis} Angular dependence of the specific
  intensity for forward (a) and backward (b) propagation obtained at
  $l_{\rm max}=21$, $g=0.98$ and $\mu_a/\mu_s=6\cdot 10^{-5}$. The
  distance to the source $z$ is assumed to be positive for forward
  propagation and negative for backward propagation.}
\end{figure}

Fig.~\ref{fig:onaxis} illustrates the specific intensity for forward
and backward propagation. Optical parameters were chosen to be close
to those of typical biological tissues in the near infrared spectral
region (see figure caption for details). The point of observation is
placed at ${\bf r}=(0,0,z)$, where $z$ is positive for forward
propagation and negative for backward propagation, and $\theta$ is
defined in both cases as the angle between the vector $\hat{\bf s}$
and the positive direction of the $z$-axis. It can be seen from the
figure that the specific intensity in the backward direction is
significantly smaller compared to that in the forward direction, even
at relatively large source-detector separations ($\vert z \vert =
6\ell^*$). It can be also seen that the angular distribution of the
specific intensity in the forward direction is more sharply peaked
than that in the backward direction. This can be explained by noticing
that backward propagation involves more scattering events than forward
propagation of the same distance.

\begin{figure}
\centerline{\psfig{file=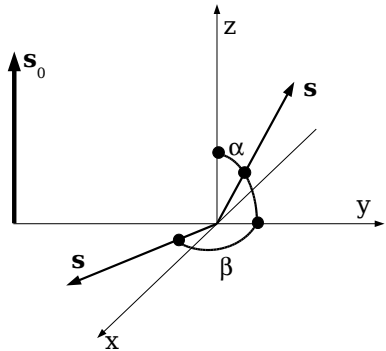,width=4in,bbllx=0bp,bblly=90bp,bburx=100bp,bbury=170bp,clip=t}}
\vspace{-3cm}
\caption{\label{fig:alpha_beta}Illustration of angles $\alpha$ and $\beta$.}
\end{figure}

\begin{figure}
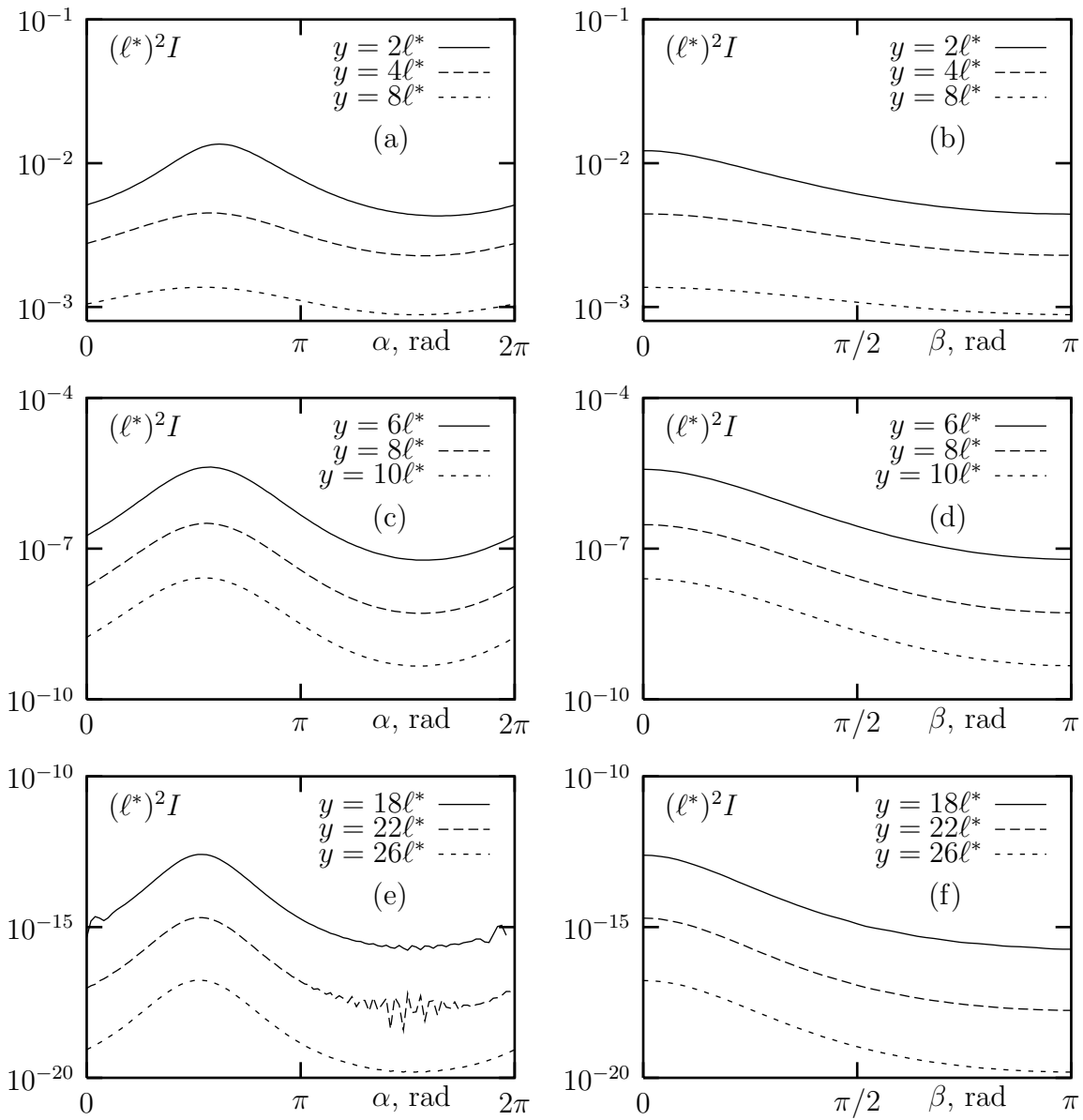

\centerline{\input{fig5a_log.tex}\hspace{-1.0cm}\input{fig5b_log.tex}}
\centerline{\input{fig6a_log.tex}\hspace{-1.0cm}\input{fig6b_log.tex}}
\centerline{\input{fig7a_log.tex}\hspace{-1.0cm}\input{fig7b_log.tex}}
\caption{\label{fig:off_axis} Angular distribution of specific
  intensity for off-axis propagation. Parameters: $g=0.98$ and
  $\mu_a/\mu_s=6\cdot 10^{-5}$ (a,b), $\mu_a/\mu_s=0.03$ (c,d),
  $\mu_a/\mu_s=0.2$ (e,f).}
\end{figure}

\begin{figure}
\centerline{
\input{fig_alph.tex}\psfig{file=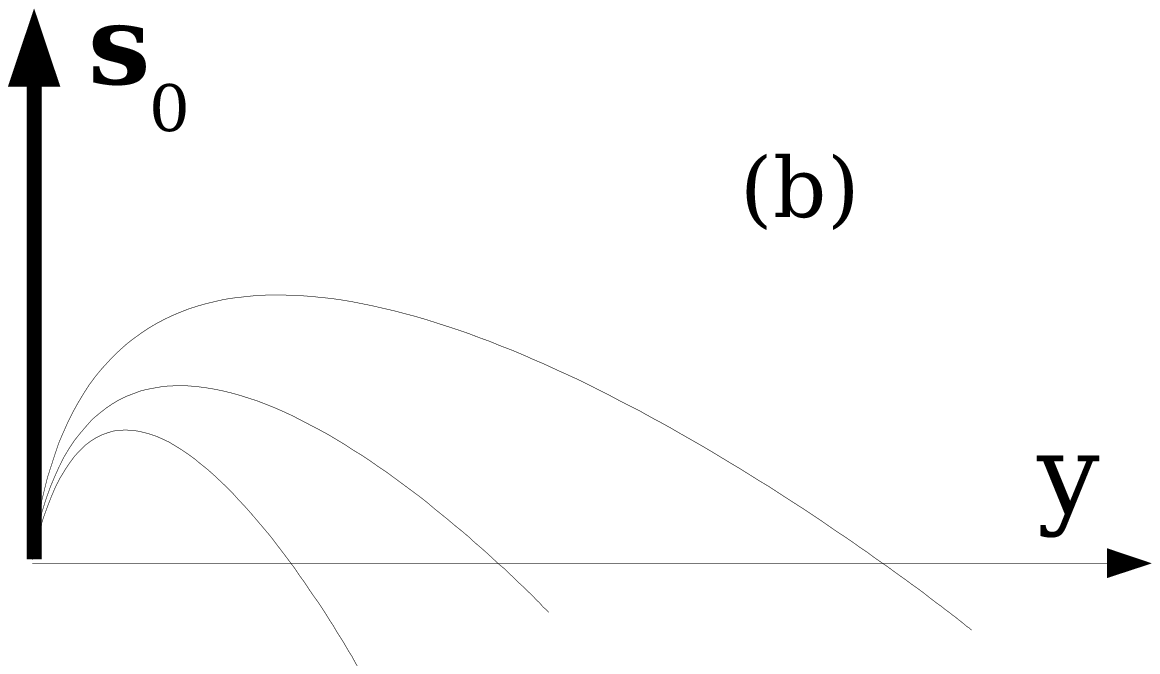,width=3in,bbllx=100bp,bblly=450bp,bburx=512bp,bbury=700bp,clip=}}
\caption{\label{fig:alpha} (a) Dependence of the position of
  maximum $\alpha_0$ on the distance to the source, $y$, for
  physiological parameters: $g=0.98$ and $\mu_a/\mu_s=6\cdot 10^{-5}$.
  (b) Schematic illustration of typical ``photon trajectories'' that
  correspond to maxima in graphs~\ref{fig:off_axis}(a,c,e).}
\end{figure}

Now we turn to the off-axis case. Here the source is still placed at
the origin and illuminates in the positive $z$ direction, while the
point of observation is placed at a point ${\bf r}=(0,y,0)$.  Below,
we show two type of graphs. In the first case, the vector $\hat{\bf
  s}$ is in the $y-z$ plane, and its orientation is characterized by
the angle $\alpha$ with respect to the positive direction of the
$z$-axis. In the second case, $\hat{\bf s}$ is in the $x-y$ plane
(perpendicular to $\hat{\bf s}_0$) and is characterized by the angle
$\beta$ with respect to the positive direction of the $y$-axis. The
angles $\alpha$ and $\beta$ (not to be confused with the Euler angles)
are illustrated in Fig.~\ref{fig:alpha_beta}. Note that $\alpha$
varies from $0$ to $2\pi$ while $\beta$ is restricted to the interval
$[0,\pi]$ due to the obvious symmetry.

In Fig.~\ref{fig:off_axis} we illustrate the specific intensity for
highly forward-peaked scattering ($g=0.98$) and the following three
different ratios of $\mu_a/\mu_s$: $6\cdot 10^{-5}$, $0.03$ and $0.2$.
Note that the corresponding ratios of $\mu_a/\mu_s^{\prime}$, where
$\mu_s^{\prime}=(1-g)\mu_s$ is the reduced scattering coefficient, are
$0.003$, $1.5$ and $10$, respectively. In the first case, the
transport mean free path is mainly determined by scattering, while in
the third case it is determined by absorption. The left column of
images (a,c,e) illustrates the angular dependence of the specific
intensity as a function of the angle $\alpha$ (vector $\hat{\bf s}$ is
in the $y-z$ plane). The oscillations visible in
Fig.~\ref{fig:off_axis}(e) is due to the non-square integrability
discussed above. However, the values of the specific intensity at the
region where the oscillations are visible are two to three orders of
magnitude smaller than those at the peak.

It is interesting to analyze the position of the maximum of the curves
in Fig.~\ref{fig:off_axis}(a,c,e), $\alpha_0$. As the distance between
the source and the detector increases, $\alpha_0$ approaches $\pi/2$.
This corresponds to a vector $\hat{\bf s}$ coinciding with the
direction from the source to detector. However, for relatively small
source-detector separations, $\alpha_0$ is larger than $\pi/2$. The
dependence of $\alpha_0$ on the source-detector separation is
illustrated in Fig.~\ref{fig:alpha}(a) for physiological parameters.
The dependence of $\alpha_0$ on the source-detector separation can be
understood at the qualitative level. Indeed, at large separations, the
angular distribution of the specific intensity is expected to be
independent of the source orientation, with the maximum attained when
$\hat{\bf s}$ is aligned with the direction from the source to the
detector.  This corresponds to $\alpha_0=\pi/2$. At smaller
separations, the ``photons'' arrive at the detector locations along
some ``typical'' (most probable) trajectories which are schematically
illustrated in Fig.~\ref{fig:alpha}(b). We assume here that $\alpha_0$
is determined by the angle at which the most probable trajectory
crosses the $y$-axis.

In Fig.~\ref{fig:off_axis}(b,d,f), the specific intensity is shown as
a function of the angle $\beta$ (vector $\hat{\bf s}$ is in the $x-y$
plane). In this case, the maximum of the curves always corresponds to
$\beta=0$, which could be also inferred from the symmetry. We note
that $I_{xy}(\beta=0)=I_{yz}(\alpha=\pi/2)$, where the lower
subscripts indicate the plane in which contains the vector $\hat{\bf
  s}$.

\begin{figure}
\centerline{\input{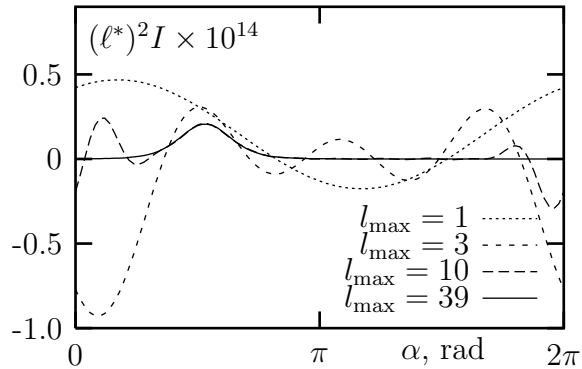}}
\caption{\label{fig:conv} Convergence of the specific intensity with
  $l_{\rm max}$ for $g=0.98$, $\mu_a/\mu_s= 0.2$ and $y=22\ell^*$.}
\end{figure}

The curves shown in Fig.~\ref{fig:off_axis} have a dynamic range of
approximately $10^3$. A dynamic range of this magnitude was obtained
due to the use of large values of $l_{\rm max}$. For smaller values of
$l_{\rm max}$, the result can be grossly inaccurate and even negative.
For example, in Fig.~\ref{fig:conv} we illustrate convergence with
$l_{\rm max}$ to one of the curves shown in
Fig.~\ref{fig:off_axis}(e).  An accurate value of specific intensity
at $\alpha\approx \pi/2$ ($\approx 10^{-3}$ relative error) was
obtained at $l_{\rm max}=39$.  Note that at $l_{\rm max}=10$, the
computed specific intensity is still grossly inaccurate.

\section{Discussion}
\label{sec:summ_disc}

The theoretical approach developed in this paper is, essentially, a
spectral approach. Spectral methods have been studied extensively for
the one-dimensional RTE~\cite{kaper_book_82}. However, in the 3D case
these methods become very difficult to use. The substantially novel
element of this paper is that we derive a usable spectral method for
the full three-dimensional RTE with an arbitrary phase function and
planar boundaries. The analytical part of the solution is of
considerable complexity. However, this complexity is traded for the
relative simplicity of the numerical part. In fact, we believe that we
have reduced the numerical part of the computations to the absolute
minimum which is allowed by the mathematical nature of the problem.

This paper is limited to consideration of spatially-independent
optical coefficients and phase functions. However, we note that the
Green's function for a macroscopically homogeneous medium is of
special interest, since it is used in linearized image reconstruction
in optical tomography~\cite{markel_04_4} and, more generally, in
nonlinear image reconstruction based on the inversion of a functional
series or the Newton-Kantorovich method~\cite{markel_03_2}. Assuming
the presence of only absorptive inhomogeneities in the medium, the
linearized kernel of the integral equation of diffusion tomography has
the form (in the slab imaging geometry)~\cite{markel_04_4}

\begin{equation}
\label{Gamma_G}
\hspace{-1cm} \Gamma({\bm \rho}_1,{\bm \rho}_2; {\bf r}) = \int G({\bm
  \rho}_1,z=0,\hat{\bf s}_1=\hat{\bf z}; {\bf r}, \hat{\bf s}) G({\bf
  r}, \hat{\bf s}; {\bm \rho}_2,z=L,\hat{\bf s}_2=\hat{\bf z}) d^2s \ ,
\end{equation}

\noindent
where ${\bm \rho}_1$ and ${\bm \rho}_2$ are the transverse coordinates
of the source and detector, respectively, located on opposite surfaces
of the slab, and $G$ is the slab Green's function with constant
absorption and scattering coefficients. One of the advantages of
solutions obtained in this paper, compared to those based on discrete
ordinates, is that the angular integral in the above formula can be
evaluated analytically.

This research was supported by the NIH under grant P41RR02305.

\appendix

\section{Calculation of the integral (\ref{chi_rs_gen_oper})}
\label{App:A}

In this Appendix we evaluate the integral (\ref{chi_rs_gen_oper}) for
different choices of $\hat{\bf z}$. Written in component form, this
integral reads

\begin{eqnarray}
\langle lm \vert \chi({\bf r};\hat{\bf z}) \vert l^{\prime}m^{\prime}
\rangle = &&  {1 \over \sqrt{\sigma_l \sigma_{l^{\prime}}}} \sum_{m_1=-l}^l
 \sum_{m_2=-l^{\prime}}^{l^{\prime}} \nonumber \\
&& \times \int {{d^3{\bf k}} \over
     {(2\pi)^3}} \exp(i{\bf k}\cdot {\bf r}) 
   \exp[-i(m-m^{\prime})\varphi_{\hat{\bf k}}] \nonumber \\
&& \times d_{mm_1}^l(\theta_{\hat{\bf k}}) d_{m^{\prime}
  m_2}^{l^{\prime}} (\theta_{\hat{\bf k}}) \sum_\mu { { \langle
       lm_1 \vert \psi_\mu \rangle \langle \psi_\mu \vert l^{\prime}m_2
       \rangle} \over {1 + ik\lambda_\mu }} \ . 
\label{Eq_A0}
\end{eqnarray}

We start with the case $\hat{\bf z}=\hat{\bf r}$. Then we have ${\bf
  k} \cdot \hat{\bf r} = kr \cos\theta_{\hat{\bf k}}$. We also notice
that $\langle l m_1 \vert \psi_\mu \rangle \langle \psi_\mu \vert
l^{\prime} m_2 \rangle \propto \delta_{m_1 m_2}$, so that the
summation over $m_1$ and $m_2$ can be replaced by summation over a
single index $M$ which runs from $-\bar{l}$ to $\bar{l}$, where
$\bar{l}=\min(l,l^{\prime})$. Then (\ref{Eq_A0}) can be rewritten
as

\begin{equation}
\hspace{-1cm} \langle lm \vert \chi({\bf r};\hat{\bf r}) \vert l^{\prime}m^{\prime}
\rangle = {1 \over \sqrt{\sigma_l \sigma_{l^{\prime}}}} \sum_{M=-\bar{l}}^{\bar{l}}
\int_0^\infty {{k^2 dk} \over
     {(2\pi)^2}}  \sum_\mu { { \langle
       l M \vert \psi_\mu \rangle \langle \psi_\mu \vert l^{\prime} M
       \rangle} \over {1 + ik\lambda_\mu }} \ \ I \ ,
\label{Eq_A1}
\end{equation}

\noindent
where $I$ is the angular part of the integral (the list of formal
arguments of $I$ is omitted):

\begin{equation}
\hspace{-1cm} I = \int {{\sin\theta_{\hat{\bf k}} d \theta_{\hat{\bf k}} d
    \varphi_{\hat{\bf k}} } \over {2\pi}}  
\exp[i(m^{\prime}-m)\varphi_{\hat{\bf k}}]
\exp(ikr \cos\theta_{\hat{\bf k}}) 
d_{mM}^l(\theta_{\hat{\bf k}}) d_{m^{\prime}
  M}^{l^{\prime}} (\theta_{\hat{\bf k}})  \ . 
\label{Eq_A2}
\end{equation}

\noindent
The integral over $\varphi_{\hat{\bf k}}$ is evaluated immediately
with the result $2\pi \delta_{m m^{\prime}}$. Integration over
$\theta_{\hat{\bf k}}$ requires expanding the exponent in the
integrand as

\begin{equation}
\label{Eq_A3}
\exp(ikr \cos\theta_{\hat{\bf k}}) =\sum_{L=0}^{\infty} i^L (2L +1)
j_L(kr) d_{00}^L(\theta_{\hat{\bf k}}) \ ,
\end{equation}

\noindent
where $j_L(x)$ are the spherical Bessel functions of the first kind,
and using the following formula (see Ref.~\cite{varshalovich_book_88},
Sect.~4.11.2, formula 8, and the symmetry properties of d-functions
given in Sect.~4.11, formula 1 of the same reference):

\begin{equation}
\int_0^{\pi} 
d_{mM}^l(\theta)
d_{mM}^{l^{\prime}}(\theta) 
d_{00}^L(\theta) 
\sin\theta d\theta = 
{{2(-1)^{m-M}}  \over {2L+1}} 
C_{l,m,l^{\prime},-m}^{L,0}
C_{l,M,l^{\prime},-M}^{L,0} \ ,
\label{Eq_A4}
\end{equation}

\noindent
where $C_{j_1m_1j_2m_2}^{j_3m_3}$ are the Clebsch-Gordan
coefficients. Taking into account that $C_{l,m,l^{\prime},-m}^{L,0}$
is nonzero only for $\vert l - l^{\prime} \vert \leq L \leq l +
l^{\prime}$, we obtain

\begin{equation}
\label{Eq_A4_I}
I = 2\delta_{m m^{\prime}} (-1)^{m-M} \sum_{L=\vert l - l^{\prime}
  \vert}^{l + l^{\prime}} i^L j_L(kr) C_{l,m,l^{\prime},-m}^{L,0}
C_{l,M,l^{\prime},-M}^{L,0} \ .
\end{equation}

\noindent
Next, we substitute this result into (\ref{Eq_A4}) and, after some
rearrangement, arrive at 

\begin{eqnarray}
\hspace{-1cm} \langle lm \vert \chi({\bf r}; \hat{\bf r}) \vert l^{\prime}
m^{\prime} \rangle = &&
{{2 \delta_{m m^{\prime}} (-1)^m } 
\over 
\sqrt{\sigma_l \sigma_{l^{\prime}}}}
\sum_{M=-\bar{l}}^{\bar{l}} (-1)^M 
\sum_{L=\vert l - l^{\prime} \vert}^{l + l^{\prime}} i^L 
C_{l,m,l^{\prime},-m}^{L,0}
C_{l,M,l^{\prime},-M}^{L,0} 
\nonumber \\ && \times 
\sum_\mu  \langle l M \vert \psi_\mu \rangle 
          \langle \psi_\mu \vert l^{\prime} M \rangle  
\int_0^\infty {{k^2 dk} \over {(2\pi)^2}}  
{{j_L(kr)} \over {1 + ik\lambda_\mu }} \ .
\label{Eq_A5}
\end{eqnarray}

\noindent
To evaluate the radial integral, we exploit the symmetry properties of
the above expression. First, we notice that
$C_{l,M,l^{\prime},-M}^{L,0} = (-1)^{l+l^{\prime}+L}
C_{l,-M,l^{\prime},M}^{L,0}$, while $\langle lM\vert \psi_\mu \rangle$
does not depend on the sign of $M$.  Thus, the addition of terms with
positive and negative values of $M$ in the above formula (for $M\neq
0$) gives zero unless $l + l^{\prime} + L$ is even.  Likewise, in the
case $M=0$, $C_{l,0,l^{\prime},0}^{L,0}=0$ unless the above sum of
indices is even. Correspondingly, the only nonzero contributions to
the sum over $L$ corresponds to $L=\vert l - l^{\prime} \vert + 2j$,
where the index $j$ runs from $0$ to $\bar{l}$. Next, we use the
symmetry property of the eigenvectors discussed in
Section~\ref{subsubsec:symmetry}. This property allows one to limit
summation over the eigenvector indices $\mu$ to only the values
corresponding to positive eigenvalues $\lambda_\mu$ while
simultaneously replacing the factor $1/(1 + i k \lambda_\mu)$ by $1/(1
+ i k \lambda_\mu) + (-1)^{l + l^{\prime}}/(1 - i k \lambda_\mu)$.
Thus, we obtain

\begin{eqnarray}
\hspace{-1cm} \langle lm \vert \chi({\bf r}; \hat{\bf r}) \vert l^{\prime}
m^{\prime} \rangle = && 
{{2 \delta_{m m^{\prime}} (-1)^m } 
\over 
\sqrt{\sigma_l \sigma_{l^{\prime}}}}
\sum_{M=-\bar{l}}^{\bar{l}} (-1)^M 
\sum_{j=0}^{\bar{l}} i^{\vert l - l^{\prime} \vert + 2j}
C_{l,m,l^{\prime},-m}^{\vert l - l^{\prime} \vert + 2j,0}
C_{l,M,l^{\prime},-M}^{\vert l - l^{\prime} \vert + 2j,0} \nonumber \\
&& \times {\sum_\mu}^{\prime}  \langle l M \vert \psi_\mu \rangle 
          \langle \psi_\mu \vert l^{\prime} M \rangle  \ \ J \ ,
\label{Eq_A6}
\end{eqnarray}

\noindent
where $J$ is the radial integral given by

\begin{equation}
J = \int_0^\infty {{k^2dk} \over {(2\pi)^2}} j_{\vert l - l^{\prime}
  \vert + 2j}(kr) 
{{1 + (-1)^{l+l^{\prime}} - ik\lambda_\mu [1 - (-1)^{l+l^{\prime}}]} 
\over
{1 + k^2 \lambda_\mu^2}} \ .
\label{Eq_A7}
\end{equation}

\noindent
The parity of the Bessel functions in the above integral is the same
as that of $l + l^{\prime}$. Therefore, the integrand is an even
function of $k$ for all values of the indices, and the integral can be
extended to $-\infty$ and calculated by residues. The result is

\begin{equation}
J = \pi i^{-(\vert l - l^{\prime} \vert + 2j) } 
\lambda_\mu^{-3} k_{\vert l - l^{\prime} \vert + 2j} (r/\lambda_\mu) \
.
\label{Eq_A8}
\end{equation}

\noindent
Upon substitution of this result into (\ref{Eq_A6}), we obtain the
formula (\ref{chi_llm_uni}).

In the case $\hat{\bf z}=\hat{\bf s}_0$ the dot product ${\bf
  k}\cdot{\bf r}$ can not be written as $kr\cos\theta_{\hat{\bf
    k}}$. Therefore, the exponent in the angular integral $I$ is
expanded as

\begin{equation}
\exp(i {\bf k}\cdot{\bf r}) = 4\pi\sum_{LM^{\prime}} i^L j_L(kr)
Y_{LM^{\prime}}(\hat{\bf k};\hat{\bf s}_0) Y_{LM^{\prime}}^*(\hat{\bf r};\hat{\bf s}_0)
\ .
\label{Eq_A9}
\end{equation}

\noindent
We further take advantage of the identity

\begin{equation}
\label{Eq_A9a}
Y_{LM^{\prime}}(\theta_{\hat{\bf k}}, \varphi_{\hat{\bf k}}) =
(-1)^{M^{\prime}} \sqrt{4\pi / (2L+1)} d_{0M^{\prime}}^l
(\theta_{\hat{\bf k}}) \exp(i M^{\prime} \varphi_{\hat{\bf k}})
\end{equation}

\noindent
to transform the angular integration to the general form
(\ref{Eq_A4_I}).  Note that azimuthal integration results in a factor
of $\delta_{M^{\prime}m}$ and thus removes summation over
$M^{\prime}$.  The final result for $I$ is

\begin{equation}
I = {{2(-1)^l \sqrt{4\pi}} \over {\sqrt{2 l^{\prime} + 1}} }
\sum_{L=\vert l - l^{\prime} \vert}^{l + l^{\prime}}
i^L j_L(kr) Y_{Lm}^*(\hat{\bf r};\hat{\bf s}_0) 
C_{l,M,L,0}^{l^{\prime},M} C_{l,m,l^{\prime},0}^{L,m} \ ,
\label{Eq_A10}
\end{equation}

\noindent
where we have also used $C_{L,M,l^{\prime}-M}^{L,0} = (-1)^{l-M}
\sqrt{(2L+1)/(2l^{\prime}+1)} C_{l,M,L,0}^{l^{\prime},M}$. The radial
integration, and the symmetry considerations explained above, remain
without change.  Substitution of (\ref{Eq_A10}) into (\ref{Eq_A1}) and
subsequent radial integration leads to the formula
(\ref{chi_llm_uni_z}).

\section{Calculation of the integral (\ref{kappa_int_kz})}
\label{App:B}

Integral (\ref{kappa_int_kz}), written in components, reads

\begin{equation}
\label{Eq_B1}
\hspace{-2cm} \langle l m \vert \kappa({\bf q};z) \vert l^{\prime} m^{\prime}
\rangle = \frac{\exp\left[-i(m - m^{\prime})\right]}{\sqrt{\sigma_l
    \sigma_{l^{\prime}}}} \sum_{M=-\bar{l}}^{\bar{l}}
{\sum}^{\prime}_n \frac{\langle l \vert \phi_n(M) \rangle \langle
  \phi_n(M) \vert l^{\prime} \rangle}{\lambda_{nM}^2} \ \ I \ .
\end{equation}

\noindent
Here $I$ is the integral over $k_z$:

\begin{equation}
\label{Eq_B2}
\hspace{-2.5cm} I = \int_{-\infty}^{\infty} 
{{d k_z} \over {2\pi}} \exp(i k_z z) d_{mM}^l(\theta)
d_{m^{\prime}M}^{l^{\prime}}(\theta) 
\frac{1 + (-1)^{l+l^{\prime}} - i\lambda_{Mn}\sqrt{q^2 + k_z^2}[1 -
  (-1)^{l+l^{\prime}}]  }
{k_z^2 + q^2 + 1/\lambda_{Mn}^2} \ ,
\end{equation}

\noindent
where we have used the notations introduced in
Section~\ref{subsubsec:block_structure} for block eigenvectors $\vert
\phi_n(M) \rangle$. The angle $\theta$ is defined by
(\ref{cos_sin_theta}) in Section~\ref{subsec:pw_dec}.  The Wigner
d-functions can be written in terms of $\cos\theta$ as

\begin{equation}
\label{Eq_B4}
\hspace{-1cm}d_{mM}^l(\theta) = \xi_{mM} Z_{mM}^l 
\left( \frac{1 - \cos\theta}{2}\right)^{\frac{\vert m - M \vert}{2}}
\left( \frac{1 + \cos\theta}{2}\right)^{\frac{\vert m + M \vert}{2}}
P_s^{(u,v)}(\cos\theta) \ ,
\end{equation}

\noindent
where $\xi_{mM}=1$ if $m\leq M$ and $\xi=(-1)^{m+M}$ if $m>M$,

\begin{equation}
\label{Eq_B5}
\hspace{-2cm} Z_{mM}^l = \sqrt{ \frac{
\left(l - \vert m - M \vert /2 - \vert m + M \vert /2 \right)! 
\left(l + \vert m + M \vert /2 - \vert m + M \vert /2 \right)! 
}{
\left(l + \vert m - M \vert /2 - \vert m + M \vert /2 \right)! 
\left(l - \vert m + M \vert /2 - \vert m + M \vert /2 \right)! 
}} \ ,
\end{equation}

\noindent
and $P_s^{(u,v)}(x)$ in expression (\ref{Eq_B4}) are Jacobi polynomials
with $s=l - \vert m - M \vert /2 - \vert m + M \vert / 2$, $u=\vert m
- M \vert$ and $v=\vert m + M \vert$.

The integrand in (\ref{Eq_B2}) is not, in general, an analytic
function of $k_z$. However, the expression for the Green's function
contains a summation over $M$. It can be shown explicitly that the
combination

\begin{equation}
\label{Eq_B6}
d_{mM}^l(\theta) d_{m^{\prime}M}^{l^{\prime}}(\theta) +
d_{m-M}^l(\theta) d_{m^{\prime}-M}^{l^{\prime}}(\theta) 
\end{equation}

\noindent
contains only even powers of the factor $\sqrt{k_z^2 + q^2}$ if
$l+l^{\prime}$ is even and only odd powers of the same factor if
$l+l^{\prime}$ is odd (a general proof of this statement is available
but omitted). Taking into account the factor $\sqrt{q^2 + k_z^2}[1 -
(-1)^{l+l^{\prime}}]$ in the right-hand side of (\ref{Eq_B2}), we
arrive at the conclusion that the integrand becomes analytic after
addition of terms with positive and negative values of $M$. Note that
the eigenvectors and eigenvalues do not depend on the sign of $M$ and
the above consideration applies to the case $M=0$. Consequently, one
can evaluate (\ref{Eq_B2}) by residues choosing a branch of the
complex-valued function $\sqrt{k_z^2 + q^2}$ arbitrarily.

The integrand of (\ref{Eq_B2}) has simple poles at $k_z = \pm i
\sqrt{q^2 + 1/\lambda_{Mn}^2}$. Taking account of these poles leads to
the following expression:

\begin{eqnarray}
I = && \frac{
[{\rm sgn}(z)]^{l + l^{\prime} + m + m^{\prime}} \lambda_{Mn} 
\exp\left[-\sqrt{1 + (q\lambda_{Mn})^2}\vert z \vert /\lambda_{Mn}
\right]
}{
\sqrt{1 + (q\lambda_{Mn})^2}
} \nonumber \\
&& \hspace{3cm} \times d_{mM}^l[i\tau(q\lambda_{Mn})]
                         d_{m^{\prime}M}^{l^{\prime}}[i\tau(q\lambda_{Mn})] \ .
\label{Eq_B7}
\end{eqnarray}

\noindent
Substitution of (\ref{Eq_B7}) into (\ref{Eq_B1}) leads to an
expression which is equivalent to (\ref{kappa_int_kz}).

We note that the integrand of (\ref{Eq_B2}) has another set of poles.
Namely, these are poles of the functions $d_{mM}^l[\theta(k_z)]$ at
$k_z=\pm iq$. These poles are of a purely geometrical nature. We have
calculated analytically the contributions of these poles to the
Green's function to the few lowest orders in $l,l^{\prime}$, and found
that they cancel each other. However, we do not have a general proof
of such cancellation to all orders. On the other hand, it is clear
that if these poles could contribute to the plane-wave decomposition
of the Green's function, the result would not satisfy the RTE since
the matrix $W$ is bounded and has no infinite eigenvalues. To confirm
the validity of the obtained analytical expression, we have computed $I$
numerically by the fourth-order Simpson rule for a model set of
parameters. Then we used this result to compute the Green's function
for the particular case $\hat{\bf s}=\hat{\bf s}_0=\hat{\bf z}$. The
result coincided with the one predicted by formula
(\ref{PW_dec_def_s=s0=z}) with machine accuracy in double precision.

\section*{References}
\bibliography{abbrevplain,article,book,tomography,local}

\end{document}